\documentclass[preprint,12pt]{casual}
\usepackage{enumitem}
\usepackage{amsmath,amsthm}
\usepackage{graphics,epsfig,subfigure}
\usepackage{algorithm,algorithmic}
\usepackage{verbatim}
\usepackage{epstopdf}
\usepackage{pgfplots}
\usepackage{lscape}
\usepackage{wrapfig}
\usepackage{rotating}
\usepackage{float}
\usepackage{lscape}
\usepackage{pdflscape}
\usepackage{multicol}
\usepackage{adjustbox}
\usepackage{nomencl}
\makenomenclature
\usepackage{savesym}
\savesymbol{pdfbookmark}
\usepackage{hyperref}
\restoresymbol{HR}{pdfbookmark}
\newcommand{\Hsection}[1]{\vspace{0.5\baselineskip}\par\noindent\textit{#1}~\textbf{---}~}
\usepackage{amssymb}
\usepackage{lineno}
\usepackage{natbib}
\begin{document}

\begin{frontmatter}

\title{Large Vocabulary Automatic Chord Estimation Using Deep Neural Nets: Design Framework, System Variations and Limitations}

\author{Junqi Deng}
\author{Yu-Kwong Kwok}
\address{Department of Electrical and Electronic Engineering, The University of Hong Kong}

\begin{abstract} 
In this paper, we propose a new system design framework for large vocabulary automatic chord estimation. Our approach is based on an integration of traditional sequence segmentation processes and deep learning chord classification techniques. We systematically explore the design space of the proposed framework for a range of parameters, namely deep neural nets, network configurations, input feature representations, segment tiling schemes, and training data sizes. Experimental results show that among the three proposed deep neural nets and a baseline model, the recurrent neural network based system has the best average chord quality accuracy that significantly outperforms the other considered models. Furthermore, our bias-variance analysis has identified a glass ceiling as a potential hindrance to future improvements of large vocabulary automatic chord estimation systems.
\end{abstract}

\end{frontmatter}

\section{Introduction}

Automatic chord estimation (ACE) has been one of the major challenges in music informatics. It can be a significant subproblem in tasks such as cover song identification \cite{bello2007audio,lee2006identifying,serra2010audio}, music structural segmentation \cite{bello2005robust}, and genre classification \cite{cheng2008automatic,perez2009genre}. It also has a critical role in problems such as audio key detection \cite{papadopoulos2012modeling,pauwels2010integrating} and downbeat estimation \cite{papadopoulos2008simultaneous,mauch2010simultaneous}.

Human chord estimation experts have developed websites such as UltimateGuitar\footnote{\url{ultimate-guitar.com}}, E-chords\footnote{\url{e-chords.com}} and many others\footnote{\url{polygonguitar.blogspot.hk}; \url{chords-haven.blogspot.hk}; \url{azchords.com}}, where the chords of millions of songs can be found. To be useful for practical purposes (e.g., song covering, busking, rehearsal, performance), those chords are often captured in great details, with a large chord vocabulary including suspensions, extensions, inversions and even alterations. As the music production rate grows, it is foreseeable that these chord services will increasingly rely on ACE technologies.

It is not uncommon to associate ACE closely with automatic speech recognition (ASR). We have witnessed advancements in ASR such as the hybrid approaches based on ``Gaussian mixture model (GMM) + hidden Markov model (HMM)'' (i.e., GMM-HMM) \cite{huang2001spoken,deng2006dynamic}, and more recently with deep learning techniques \cite{deng2014deep,yu2011deep}. Naturally, one would consider that ASR solutions can also be applied to ACE with similar success \cite{sheh2003chord}. However, while ASR requires the algorithm to output a sequence of words, ACE requires the algorithm to output a \textit{time-aligned sequence} of chords that matches the chord onsets in the input sequence. Therefore, unlike ASR, ACE requires both segmentation and classification techniques.

Most ACE systems do have similar designs as ASR systems. However, by following the ASR tradition, they perform segmentation and classification in one single pass, overlooking a possible ``divide and conquer'' strategy towards the two tasks. The problem scenario in ACE differs from ASR in that chords are usually segmented rhythmically.

\subsection{Brief Overview of ACE Systems}

The ACE problem has been studied for around two decades. The very first approach \cite{fujishima1999realtime} takes a sequence of pitch class profiles (PCP), or chromagram (a sequence of salience vectors for each of the twelve pitch classes) \cite{wakefield1999mathematical}, as the audio feature, and decodes the chord sequence using a template based mean-smoothing method. Several subsequent approaches \cite{sheh2003chord,bello2005robust} replace this method by GMM-HMM based probabilistic models. Motivated by the success of GMM-HMM, many machine learning based ACE methods that derive the GMM-HMM parameters by data have emerged \cite{weller2009structured,khadkevich2011time,ni2012end,cho2013mirex}.

Differences in implementation details notwithstanding, most traditional ACE systems are based on a similar architecture, in which the chromagram is extracted as feature, and then a GMM-HMM-like (or conditional random field \cite{burgoyne2007cross,weller2009structured}, Markov logic network \cite{papadopoulos2012modeling}) is used to decode the chromagram for a chord sequence. Amongst the most prominent of those systems are the dynamic-Bayesian-network based musical probabilistic system \cite{mauch2010mirex}, and its HMM version - Chordino \cite{mauchsimple}.

Recently, deep learning based approaches have started to emerge in ACE.  To very briefly recap, there are: 1. a convolution neural network (CNN) based system \cite{humphrey2012rethinking}, that locally processes each input frame using a CNN and then globally post-processes the CNN outputs with median-filtering for a chord sequence; 2. a hybrid fully-connected neural network (FCNN) + recurrent neural network (RNN) system \cite{boulanger2013audio}, that locally computes posterior chord probabilities for each frame using an FCNN, and globally classifies a chord sequence using an RNN; 3. a hybrid deep belief network (DBN) + RNN system \cite{sigtia2015audio}, and a hybrid DBN + HMM system \cite{zhou2015chord}, both of which are variants of the previous FCNN+RNN system. All of them have shown comparable or better results than the state-of-the-art in metrics with \textit{major} and \textit{minor} triads.

\subsection{ACE Evaluations}\label{sec:aceeval}
According to the reports\footnote{\url{http://www.music-ir.org/mirex/wiki/MIREX\_HOME}} of the annual ``music information retrieval evaluation exchange'' \cite{downie2008music} (MIREX), ACE evaluations before 2013 had mainly focused on ``\textit{MajMin}'' vocabulary, which contains 12 \textit{major} chords, 12 \textit{minor} chords and an ``NC'' chord (representing anything that cannot be described by ``chords''). Unfortunately, this vocabulary is far from covering all chords in practice.

To form a more complete evaluation, a necessary first step is to incorporate chord inversions and seventh chords into the vocabulary. Since 2013, there are 4 evaluation vocabularies in MIREX ACE: ``\textit{MajMinBass}'', which contains ``\textit{MajMin}'' and their first and second inversions; ``\textit{Sevenths}'', which contains \textit{7}, \textit{min7} and \textit{maj7} beyond ``\textit{MajMin}''; and ``\textit{SeventhsBass}'', which contains ``\textit{Sevenths}'' and all of their inversions. An ACE system can be designed to support any vocabulary, and the MIREX ACE evaluation tool \cite{pauwels2013evaluating, raffel2014mir_eval} will try to perform necessary chord mappings based on the evaluation vocabulary being used.

\subsection{Chord Inversions}

Of all the systems submitted to MIREX ACE after the new evaluation standard, only one supports chord inversions \cite{burgoyne2014comparative}. Due to the dominating population of root position chords (mainly root position triads), systems that do not support inversions could achieve relatively higher scores than those that support inversions \cite{deng2016chord} under \textit{SeventhsBass} evaluation. This is because a chord's inversion is easy to be confused with its root positions. Since the vast majority of chords are in root positions (the evaluation datasets contain mainly pop and rock music), the non-supportability of inversions makes such confusions only possible in one direction (i.e., inversions misclassified as root positions) and thus much less likely than those of the other direction (i.e., root positions misclassified as inversions).

\begin{figure}[h!]
	\centering
	\includegraphics[width=0.6\columnwidth]{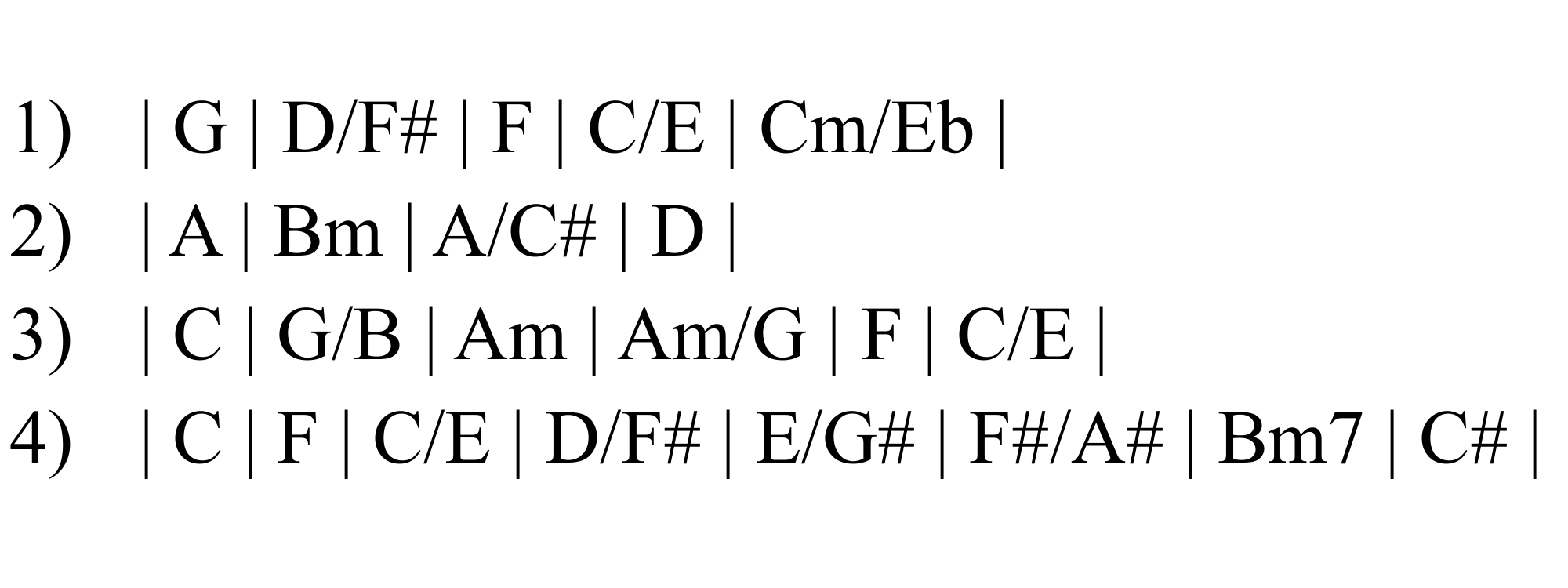}
	\caption{Four chord progressions that contain bass line continuations which require chord inversions (those with a ``/'' mark). In all these sequences, the basses are either walking downward or upward. Progressions like 1, 2 and 3 are very popular among pop/rock. Progression 4 induces a key shift from C \textit{major} to F\# \textit{minor}.}
	\label{fig:wdwu}
\end{figure}

Musically speaking, an ACE system should distinguish root positions from inversions because their sound qualities are different in many musical contexts. For example, referring to the chord progressions in Figure~\ref{fig:wdwu}, if a system does not support inversions, it breaks bass line continuations and thus alters the harmonies. As the ultimate goal of ACE is to implement music intelligence to match human experts' performance on chord recognition, the supportability of a more sophisticated vocabulary with chord inversions should be considered given this goal. Consequently, in this paper, all proposed systems support exactly the ``\textit{SeventhsBass}'' vocabulary and they all undergo the ``\textit{SeventhsBass}'' evaluation. Therefore, in this context, the \textit{large vocabulary} is referring to the set of ``\textit{SeventhsBass}'' chords (containing \textit{maj}, \textit{maj/3}, \textit{maj/5}, \textit{min}, \textit{min/b3}, \textit{min/5}, \textit{maj7}, \textit{maj7/3}, \textit{maj7/5}, \textit{maj7/7}, \textit{7}, \textit{7/3}, \textit{7/5}, \textit{7/b7}, \textit{min7}, \textit{min7/b3}, \textit{min7/5}, \textit{min7/b7}, and \textit{N}, totally 18 + 1 types) with 19 chord types in total.

\subsection{Contributions and Findings}
Thus far, we have identified several research gaps. Firstly, the existing works have not considered a fundamental difference between ASR and ACE in regards to segmentation, which may lead to a possible design that considers segmentation and classification as two separate tasks. Secondly, the support for large vocabulary has been largely overlooked, particularly the support for chord inversions. Note that chord inversion is a crucial ingredient for pop and rock music, which is the primary focus of ACE research.

As a prelude to this work, recently we proposed a hybrid ``chromagram extraction + deep neural network'' system that classifies chords based on a pre-segmented sequence [reference made anonymous on purpose]. In this paper, we generalize this system as a deep learning based large vocabulary ACE (LVACE) design framework. The main contributions and findings are as follows:
\begin{itemize}
\item we propose an LVACE system design framework using a combination of pre-segmentation techniques and deep neural nets;

\item we find that amongst all the investigated deep neural nets, the recurrent neural network performs the best in overall chord symbol recall, and significantly better than other systems in average chord quality accuracy;

\item we find that there is a glass ceiling that potentially hinders the progress of the current LVACE research.
\end{itemize}
The rest of the paper is organized as follows: Section~\ref{sec:sysframe} overviews the LVACE system framework; Sections~\ref{sec:exper} and~\ref{sec:res} describes the system implementation under the proposed framework and explores the variations under a wide range of parameters; Finally, Section~\ref{sec:concln} concludes the paper and discusses possible future LVACE research directions.

\section{System Framework} \label{sec:sysframe}

\begin{figure}[h!]
	\centering
	\includegraphics[width=0.6\columnwidth]{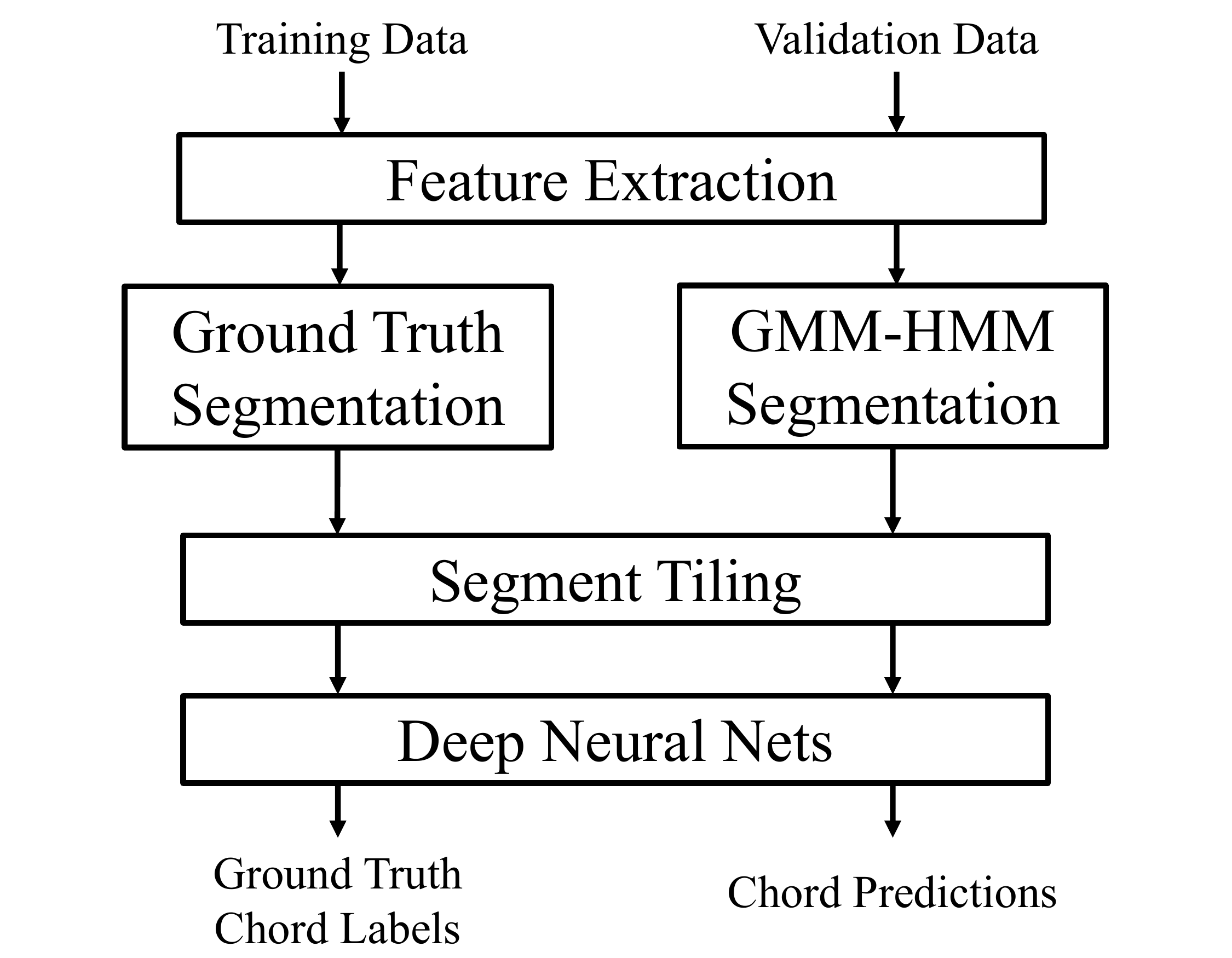}
	\caption{The LVACE system framework. The LVACE system framework. For training, each data sample first goes through a feature extraction process, and then segmented by ground truth labels. After segment tiling, the data samples are used to train a deep neural nets. The validation process shares the same modules with training, except for segmentation, where a GMM-HMM method is applied.}
	\label{fig:sysover}
\end{figure}

Figure~\ref{fig:sysover} depicts the LVACE system framework in our study. The workflow is as follows:
\begin{itemize}
	\item Feature extraction: both training and validation data share the same feature extraction module; features are extracted from each input track using a method similar to the one employed in the Chordino system~\cite{mauch2010automatic} (to be elaborated in Section~\ref{sec:fe}).
	\item Segmentation: 1. for training, the feature sequence is segmented by the ground truth annotations; 2. for validation, the feature sequence is segmented using a GMM-HMM process (to be discussed in Section~\ref{sec:sg}).
	\item Segment tiling: each feature segment is tiled into a fixed number of sub-segments (see Section~\ref{sec:seg-tile}).
	\item Deep neural nets: 1. for training, the segments and their chord labels are used to train the deep neural nets (will be described in Section~\ref{sec:dlmodel}); 2. for validation, the trained neural network is used to predict chord labels.
\end{itemize}

\subsection{Feature Extraction} \label{sec:fe}

Feature extraction starts by resampling the raw audio input at 11025 Hz, which is followed by a short-time-Fourier-transform (STFT, 4096-point Hamming window, 512-point hop size). It then proceeds to transform the linear-frequency spectrogram (2049-bin) to log-frequency spectrogram (252-bin, three bins per semitone ranging from MIDI note 21 to 104) using two cosine interpolation kernels \cite{mauch2010automatic}. The output at this step is a log-frequency-spectrogram, or log-spectrogram, $Y_{k,m}$, where $k$ is the index of frequency bins, and $m$ is the index of time frames. We denote the total number of frames as $M$, and the total number of bins in each spectrum as $K$ (in this context $K$ = 252).

The amount of deviation from standard tuning is estimated using the algorithm in \cite{mauch2010simultaneous}, where the amount of detuning is estimated as:
\begin{equation}\label{eq:2-tuning}
\delta = { {wrap(-\varphi-{2\pi}/3)} \over {2\pi} },
\end{equation}
where $wrap$ is a function wrapping its input to $[-\pi,\pi)$. $\varphi$ is the phase angle at $2\pi/3$ of the discrete-Fourier-transform (DFT) of ${\sum_m Y_{k,m}} / M$. The tuning frequency $\tau$ is then computed as:
\begin{equation}
\tau=440\cdot2^{\delta/12},
\end{equation}
and the original tuning is updated by interpolating the original spectrogram $Y_{k,\cdot}$ at $Y_{{k+p},\cdot}$, where:
\begin{equation}
p = (\log(\tau / 440) / \log(2)) \times 36.
\end{equation}
The ``36'' indicates that there are 36 bins per octave (3 bins per semitone) in $Y_{k,\cdot}$. The updated log-spectrogram will be referred to as ``notegram'' in the following.

To enhance harmonic content and attenuate background noise, a standardization process is performed along the frequency axis:
\begin{equation}
Y^{STD}_{k,\cdot} = 
\begin{cases}
{{Y_{k,\cdot} - \mu_{k,\cdot}} \over \sigma_{k,\cdot}}, \text{\quad\quad\quad if } Y_{k,\cdot} > \mu_{k,\cdot}\\
0 \quad\quad\quad\quad\quad\quad\quad\text{otherwise,}
\end{cases}
\end{equation}
where $\mu_{k,\cdot}$ and $\sigma_{k,\cdot}$ are the mean and standard deviation of a half-octave window centered at $Y_{k,\cdot}$, respectively. $Y$ is then updated by $Y^{STD}$.

This is followed by a non-negative least square (NNLS) method to extract a sequence of note activation patterns \cite{mauch2010approximate} from the log-spectrogram. Concretely, assume each note activation pattern has $L$ bins (in this context $L$ = 84), the log-spectrum $Y_{\cdot,m}$ can be expressed as:
\begin{equation} \label{eq:2-nnls}
Y_{\cdot,m} \approx EX_{\cdot,m},
\end{equation}
where $E$ is a $K \times L$ matrix, which is a dictionary of note harmonic series profiles, and $X_{\cdot,m}$ is the note activation pattern to be fitted by the algorithm. The $k_{th}$ entry of $E$ is a geometrically declining overtone series \cite{gomez2006tonal_b} of length $L$:
\begin{equation}
a_l=s^{l-1},\,s\in(0,1),
\end{equation}
where $l$ indicates the $l^{th}$ upper partials of tone $k$ of the original frequency axis, and $s$ is a declining factor controlling the steepness of the partials' envelope. A large $s$ means a slower decline. Normally $s$ is within $[0.6, 0.9]$.  The NNLS algorithm \cite{lawson1995solving} is used to find out an $X_{\cdot,m}$ that minimizes the difference between $Y_{\cdot,m}$ and $EX_{\cdot,m}$. The output of this process is usually called NNLS chromagram, or NNLS matrix (84-bin, 1 bin per semitone).

\begin{figure}[h!]
	\centering
	\includegraphics[width=0.6\columnwidth]{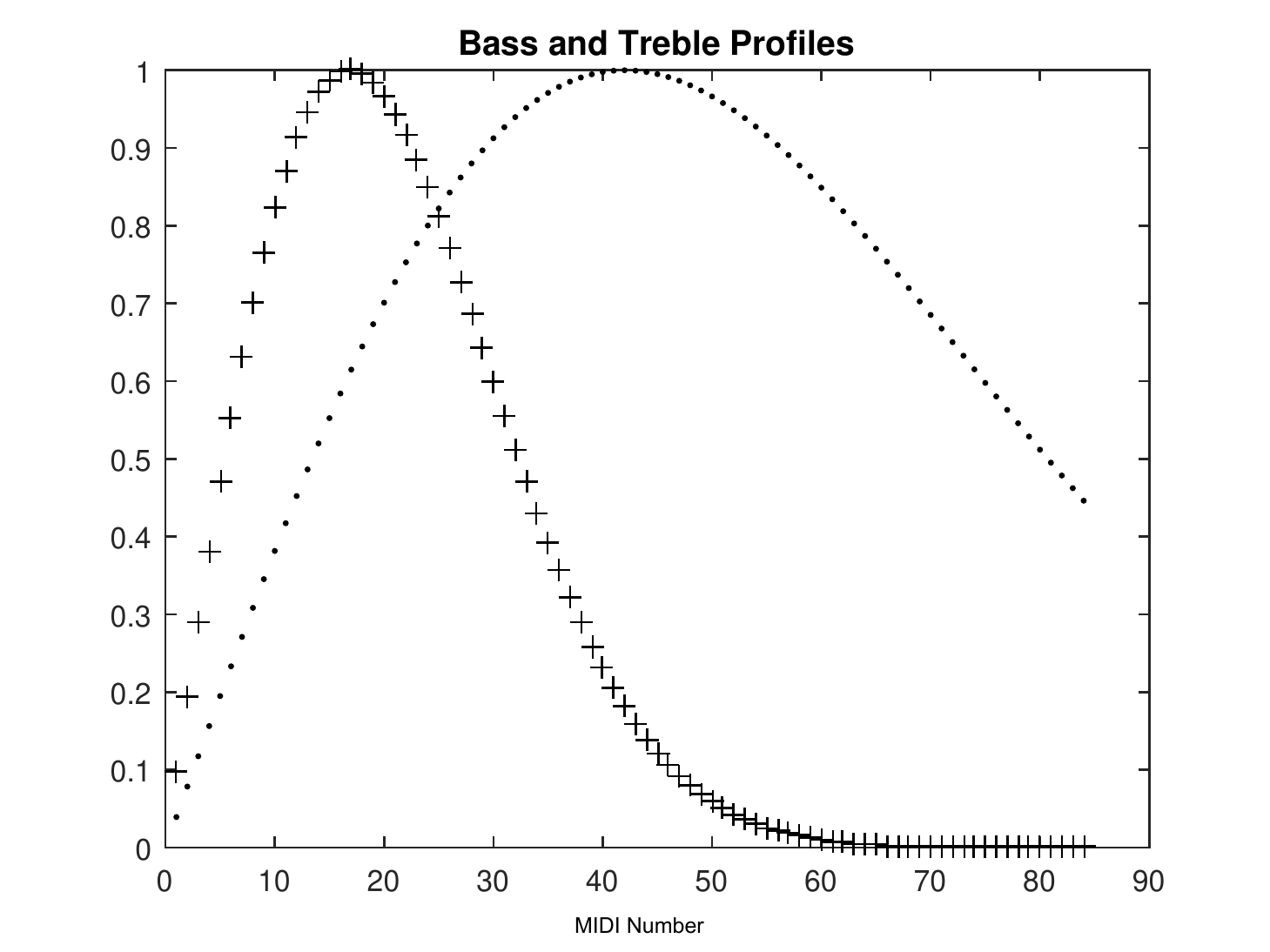}
	\caption{The bass (+) and treble (.) profiles. They are both computed in the shape of Rayleigh distributions with scale parameters 16.8 (for bass) and 42 (for treble) respectively. The horizontal axis and vertical axis represents the MIDI pitch numbers, and normalized profile amplitudes from 0 to 1, respectively.}
	\label{fig:btprofile}
\end{figure}

The feature dimension is further reduced before the segmentation process. Particularly, each NNLS chroma is weighted by the bass and treble profiles depicted in Figure~\ref{fig:btprofile}. After that the saliences of each pitch class are added together, resulting in a 24-bin bass-treble chromagram. 
Each column of the bass-treble chromagram is then $L_\infty$ normalized, so that each bin of each chroma is within the range of [0,1].

\begin{figure}[h!]
	\centering
	\includegraphics[width=\columnwidth]{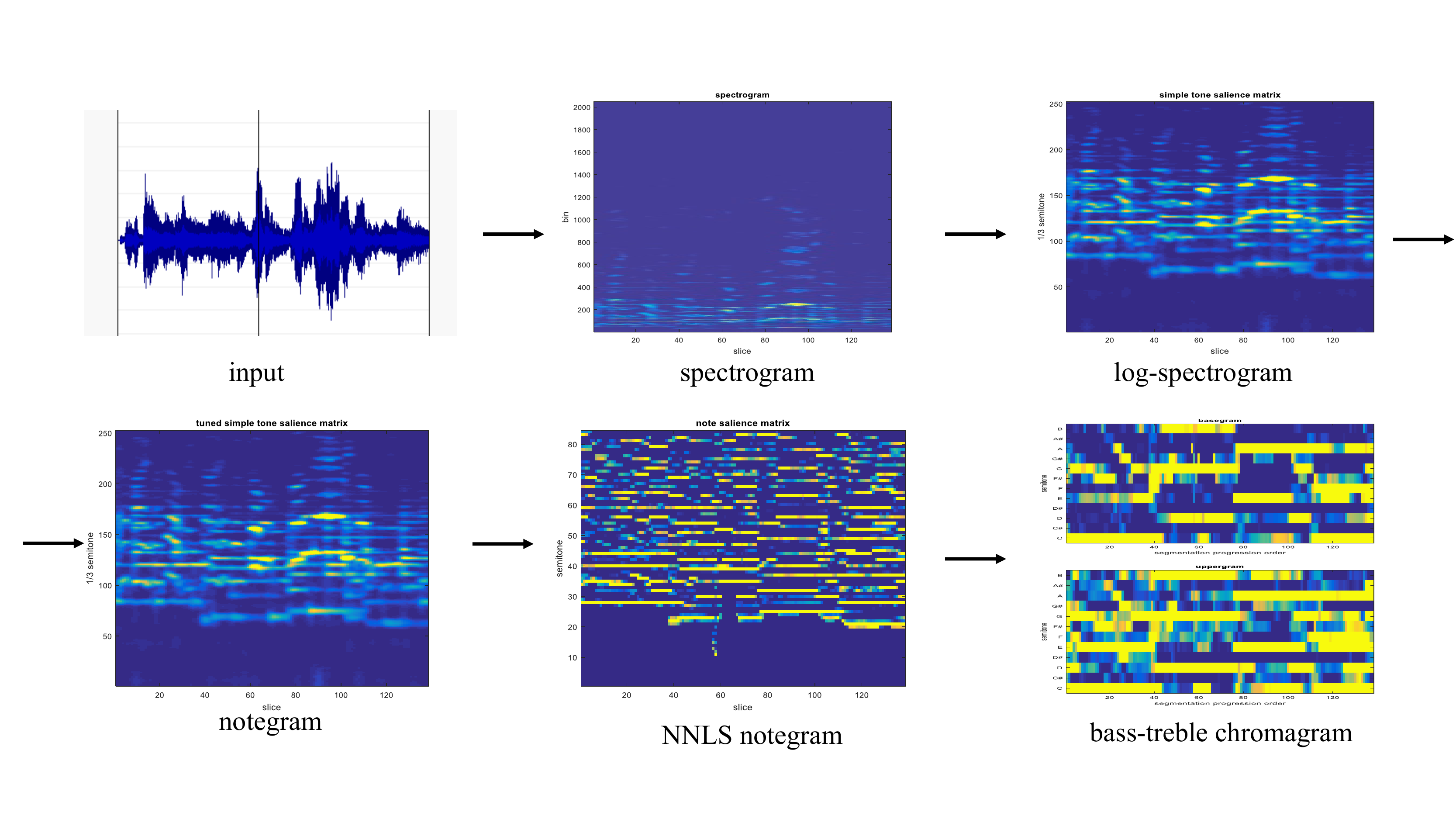}
	\caption{Information flow of feature extraction process}
	\label{fig:cdflow}
\end{figure}

Table~\ref{tab:felevels} provides a summary of different levels of representations generated by this feature extraction process. In this paper, we mainly make use of two features from the above process: the notegram and the bass-treble chromagram (simply refered to as ``chromagram'' in the following). In the experiment section, we sometimes also refer to notegram as ``-ns'', and chromagram as ``-ch''. Figure~\ref{fig:cdflow} summarizes the full information flow of the above feature extraction and segmentation process using the first line of {\it Let it be} as input.
\begin{table}[h!]
	\centering
	\footnotesize
	\begin{tabular}{|c|c|c|} \hline
		Process & Output level & Bins \\ \hline
		STFT & spectrogram & 2049 \\ \hline
		Linear-Log Mapping & log-spectrogram & 252  \\ \hline
		Tuning & notegram & 252 \\ \hline
		NNLS & NNLS notegram & 84  \\ \hline
		Bass-treble Profiling & (bass-treble) chromagram & 24 \\ \hline
	\end{tabular}
	\caption{Different levels of feature representations.}
	\label{tab:felevels}
\end{table}

\subsection{Segmentation} \label{sec:sg}

The segmentation process is implemented using a GMM-HMM, which is characterized as follows:
\begin{itemize}
	\item The hidden node models the categorical states of chords. In the \textit{SeventhsBass} implementation, there are totally 217 states (1 state per chord), where the 217 is found by multiplying the number of chord types (18) with the number of chord roots (12) and adding the number of ``N.C.'' chord types (1).
	
	\item The observable node represents a chroma. It is a 24-dimension Gaussian node connecting to the bass-treble chromagram.
	
	\item The emission of each hidden state is a 24-dimension Gaussian distribution with parameters specified in Table~\ref{tab:gaussian}. These parameters assign different Gaussians to different pitch classes according to their roles in a chord.
	
	\item The transition matrix has heavy uniform self-transition weights, which are 99.99 times of the uniform non-self-transition weight.
	
	\item The prior probabilities are uniformly distributed.
\end{itemize}
\begin{table}[h!]
	\centering
	\footnotesize
	\begin{tabular}{|c|c|c|} \hline
		& $\mu$ & $\sigma^2$ \\ \hline
		Bass - chord bass & 1 & 0.1 \\ \hline
		Bass - not chord bass and is chord note & 1 & 0.5  \\ \hline
		Bass - neither chord bass nor chord note & 0 & 0.1 \\ \hline
		Treble - chord note & 1 & 0.2  \\ \hline
		Treble - not chord note & 0 & 0.2 \\ \hline
		No Chord (for all notes)  & 1 & 0.2  \\ \hline
	\end{tabular}
	\caption{GMM-HMM segmentation settings. Different parameters are assigned to different note degree within a chord.}
	\label{tab:gaussian}
\end{table}

\subsection{Segment Tiling} \label{sec:seg-tile}

The segment tiling process is introduced to equalize the length of every segment, so as to enable neural networks with fixed-length input. This process divides a segment into $N$ equal-sized sub-segments, and takes a frame-average within each sub-segment, resulting in an $N$-frame segment (referred to as $N$seg in the following, where $N$ is a variable). If the original number of frames is not divisible by $N$, the last frame is extended to make it divisible, i.e. this process turns a segment with a variable number of frames into a segment with a fixed number of $N$ frames.

\subsection{Deep Neural Nets} \label{sec:dlmodel}

Each $N$seg will be classified as a chord label through a deep neural net. There are three types of deep neural nets considered here:  fully-connected neural network (FCNN), deep belief network, and recurrent neural network.

\subsubsection{Fully-connected Neural Network}

The FCNN is a vanilla neural network with the most basic settings. It is a feedforward neural network, and each layer is fully-connected to the next layer. It applies rectified linear units (ReLUs) as hidden layer activations. It is trained via stochastic gradient descent with back-propagation.

\subsubsection{Deep Belief Network}

The DBN is implemented based on the FCNN. Its multiple hidden layers have sigmoid activations instead of ReLUs.

In the pre-training phase, every pair of adjacent layers (except for the output layer) are trained one pair at a time as restricted Boltzmann machines (RBMs) \cite{hinton2006fast}. In our implementation, the RBM formed by the input layer and the first hidden layer is a Gaussian-Bernoulli RBM, because the input $N$seg feature contains real numbers. The RBMs formed by the hidden layer pairs are Bernoulli-Bernoulli RBMs, because each neuron is stochastic binary \cite{hinton2006reducing}.

In the fine-tuning phase, the network is regarded as a feedforward neural network and trained via stochastic gradient descent with back-propagation.

\subsubsection{Recurrent Neural Network}

\begin{figure}[h!]
	\centering
	\includegraphics[width=0.6\columnwidth]{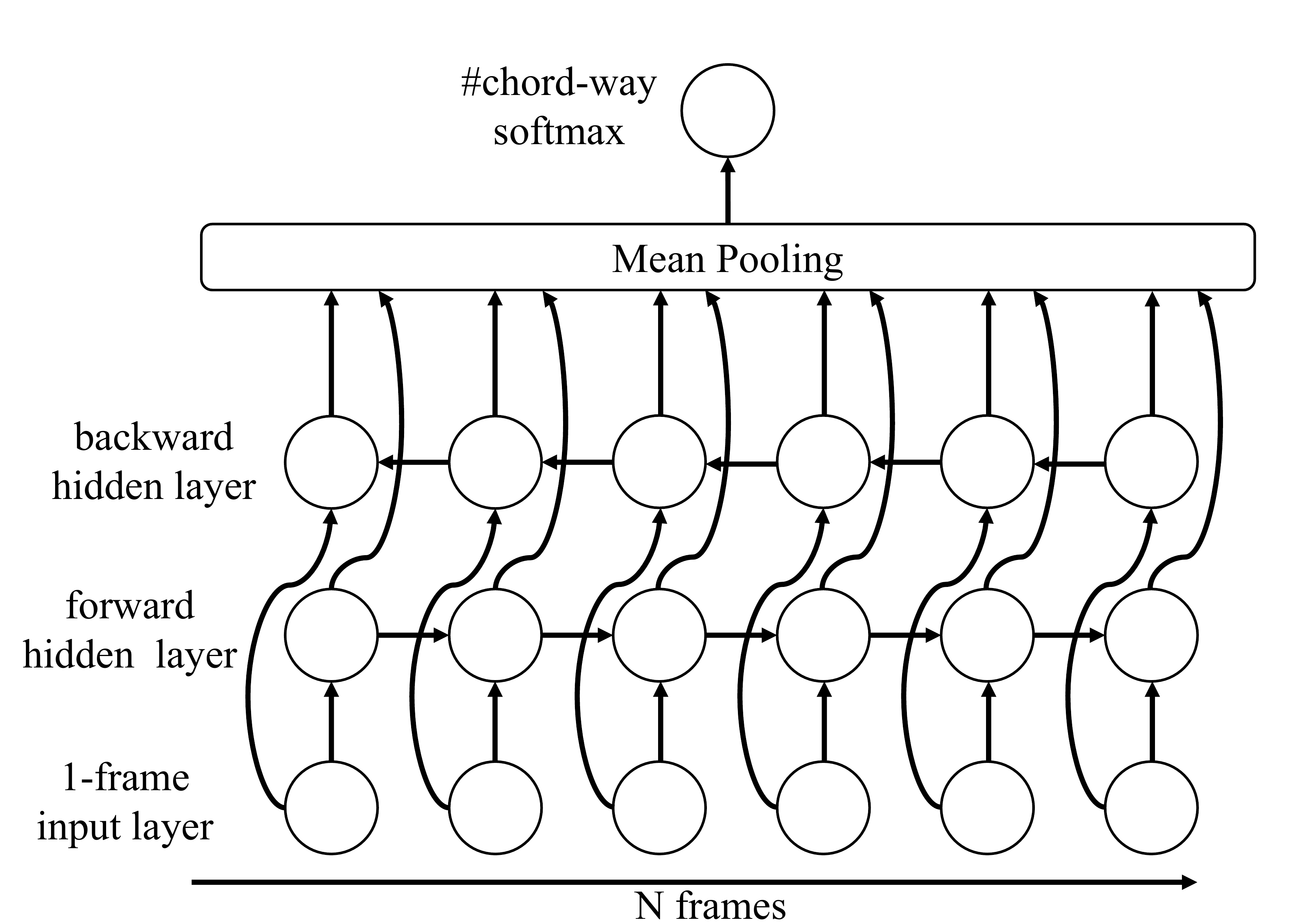}
	\caption{The bidirectional recurrent neural network architecture used in our system. Both hidden layers employ LSTM units in place of normal logistic units. The RNN is expanded to $N$ frames, with mean pooling to summarize results.}
	\label{fig:brnn}
\end{figure}

\begin{figure}[h!]
	\centering
	\includegraphics[width=0.3\columnwidth]{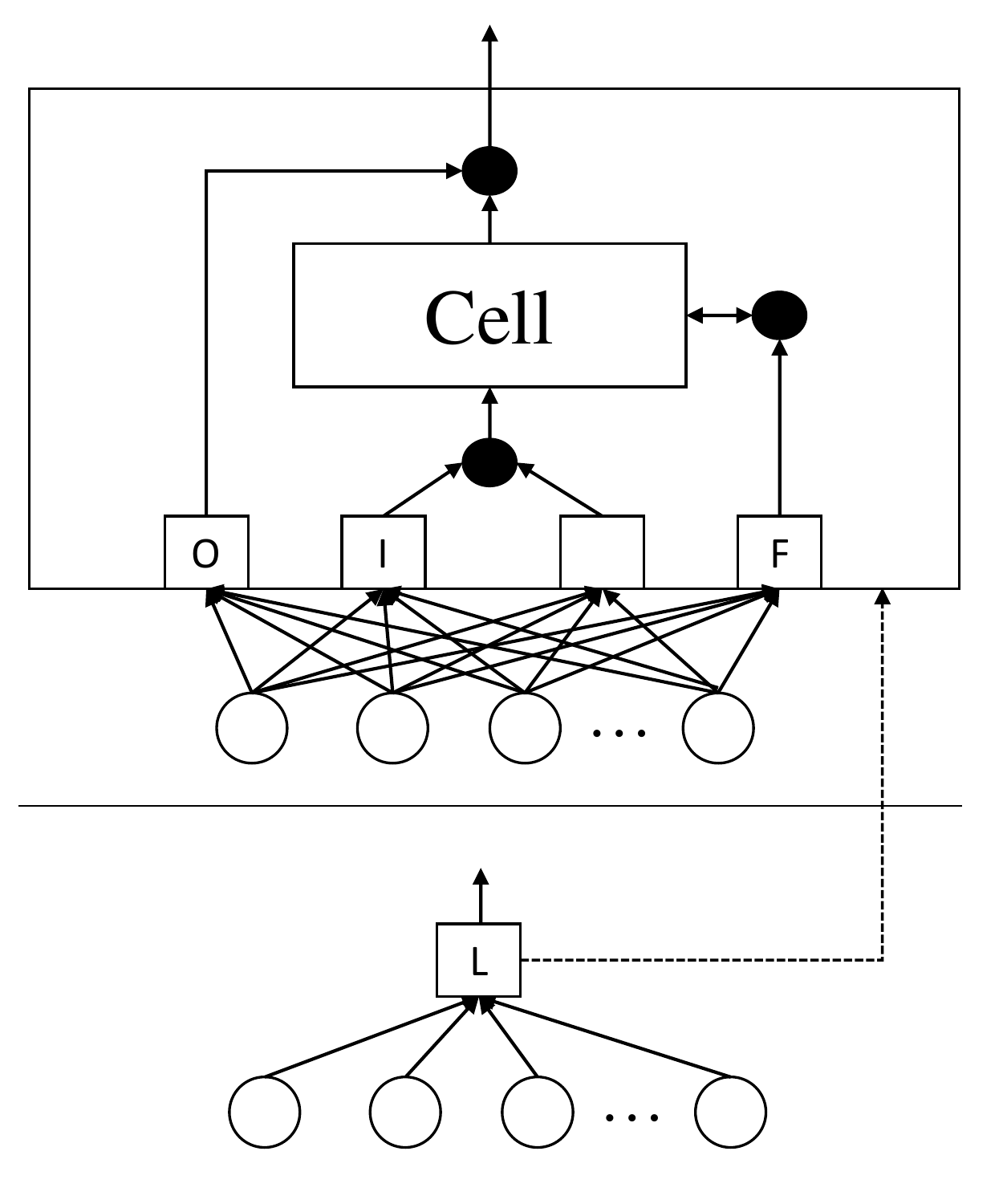}
	\caption{The structure of a long short-term memory unit}
	\label{fig:lstm}
\end{figure}

The RNN (Figure~\ref{fig:brnn}) is configured with bidirectional long-short-term-memory (LSTM) hidden units \cite{hochreiter1997long} (BLSTM-RNN). Figure~\ref{fig:lstm} shows the structure of a basic LSTM unit \cite{graves2012supervised} in the context of a simple logistic regression. The input data path has 4 identical copies for input gate $I$, output gate $O$, forget gate $F$ and the input port. Each gate or port will activate an output between 0 and 1 according to its input and activation function. The input gate activation is multiplied with the input port activation to become an input value to the LSTM cell. The forget gate activation is multiplied with the cell value from the previous time step to become another input to the cell. The current cell value is determined by the sum of the these two cell inputs. The output of the unit is given by the multiplication of the output gate activation and the current cell value. Note that in some configuration the cell value can be fed back into the gates.

LSTM can relieve the gradient vanishing problem for long sequence training \cite{bengio2009learning}. In our LSTM implementation, all gates employ sigmoid activations, while both the cell and output neuron use hyperbolic tangent activations. For a fixed-length $N$seg input, the RNN is unrolled into $N$ slices, each handling one input frame. A mean pooling operation is added before the output layer to summarize the LSTM outputs.

\section{Experiments} \label{sec:exper}

In this section, we describe a systematic approach to explore and evaluate different system variants of the LVACE framework. We first introduce the datasets, then elaborate the experimental setup, and finally discuss the training and cross-validation (CV) process.

\subsection{Datasets}

For training/cross-validation, we use six datasets of 546 tracks in total. They contain both eastern and western pop/rock songs. They are:
\begin{itemize}
	\item 29 tracks from the JayChou dataset (JayChou29, or J) \cite{dengmirex};
	\item 20 tracks from a Chinese pop song dataset (CNPop20, or C) \footnote{\url{http://www.tangkk.net/label/}};
	\item 26 tracks from the Carole King + Queen dataset (K) dataset \footnote{\url{http://isophonics.net/datasets}};
	\item 191 songs from the USPop dataset (U) \footnote{\url{https://github.com/tmc323/Chord-Annotations}};
	\item 100 tracks from the RWC dataset (R) \footnote{\url{https://staff.aist.go.jp/m.goto/RWC-MDB/}};
	\item 180 tracks from the TheBeatles180 (B) dataset \cite{harte2010towards}. 
\end{itemize}
The datasets are notated by their letter codes. For example, the combination of all datasets is denoted as ``CJKURB''.

Both chromagram (-ch) and notegram (-ns) are extracted from each track. Both of them can be transposed to all 12 different keys by circular pitch shifting (for -ch) or pitch shifting with zero paddings (for -ns). For example, a piece of treble chromagram in key of $C$ can be represented as:
\begin{equation}
PCP_C = (C',C\#', D', D\#', E', F', F\#', G', G\#', A', A\#', B'),
\end{equation}
where $X'$ stands for the salience of pitch class $X$. It can be circularly shifted to represent an equivalent PCP in other keys, such as key of $D$:
\begin{equation}
PCP_D = ( D', D\#', E', F', F\#', G', G\#', A', A\#', B', C',C\#').
\end{equation}
As for notegram, although we have pitch saliences instead of pitch class saliences, the same ``pitch shifting'' ideas can still be applied, given that the out-shifted saliences are filled by zeros.

In practice, the original key is considered as a pivot, and the features are circularly shifted to all 12 keys (the amount of transpositions ranging from -6 to 5 semitones). Adjusting the ground truth chord labels accordingly, this results in a 12-time data augmentation, which helps in reducing over-fitting \cite{cho2014improved,humphrey2015exploration}.

\subsection{Experimental Setup}
Under the proposed LVACE framework, possible design choices are:
\begin{itemize}
	\item type of deep neural nets
	\item depth and width of hidden layers (network configurations)
	\item number of frames in segment tiling
	\item input feature representations
	\item amount of training data
\end{itemize}

Our study is based on the settings depicted in Table~\ref{tab:varexplore}. For naming conventions: a combination of layer width and depth is denoted as [$width$*$depth$], such as [800*2]; a segmentation tiling scheme is denoted as $N$seg, such as 6seg; a point in this six dimensional hyper-parameter space is denoted by concatenating each parameter with ``-'', such as FCNN-6seg-[800*2]-ch-JKU. The space can be explored by parameter sweeping along a given dimension. Particularly, we will first explore along the {\it layer width} and {\it layer depth}. We then explore the \textit{segment tiling} scheme with fixed {\it layer width} and {\it layer depth}. Following the same strategy, we explore all factors in Table~\ref{tab:varexplore}. This process does not search the whole hyper-parameter space. However, it could gain us some insights of the proposed LVACE framework and produce some good system variants as well.
\begin{table}[h!]
	\centering
	\footnotesize
	\begin{tabular}{|c|c|c|} \hline
		Dimension & Variation \\ \hline
		neural net & FCNN; DBN; BLSTM-RNN \\ \hline
		segment tiling & 1; 2; 3; 6; 9; 12 (seg) \\ \hline
		layer depth & 2; 3; 4 \\ \hline
		layer width & 500; 800; 1000 \\ \hline
		input feature & notegram (-ns); chromagram (-ch) \\ \hline
		amount of training data & JK; JKU; JKUR; JKURB \\ \hline
	\end{tabular}
	\caption{Variations considered in this study}
	\label{tab:varexplore}
\end{table}

In this context we regard a ``model'' as a crossing point of all dimensions in Table~\ref{tab:varexplore}, including training data size. We regard a ``system'' as a full implementation of the LVACE framework, including the feature extraction, segmentation and deep neural nets. However, since all models share the same feature extraction and segmentation processes, we sometimes use the terms ``model'' and ``system'' interchangeably.

\subsection{Training and Cross-validation}\label{sec:traintest}

The following training procedures are applied throughout the experiments: 
\begin{itemize}
	\item Each FCNN is trained using mini-batch stochastic gradient descent, and regularized with dropout \cite{srivastava2014dropout} and early-stopping \cite{prechelt1998automatic}. 
	\item Each DBN is pre-trained using contrastive-divergence \cite{hinton2006fast} (CD-10), for 30 epochs with a learning rate of 0.001. It is fine-tuned using the FCNN's training procedure.
	\item Each BLSTM-RNN is trained using an Adadelta optimizer \cite{zeiler2012adadelta}, regularized with dropout and early-stopping. 
	\item All mini-batch stochastic gradient descents use a learning rate of 0.01 and a batch size of 100. 
	\item All early-stopping criteria are monitored using the validation error of the CNPop20 dataset, which is not in any training set. The model with the lowest validation loss will be saved, and if the current validation loss is 0.996 times smaller than the lowest one, the early-stopping patience will increase by the value of the current number of iterations. Training stops when the patience is less than the current number of iterations.
	\item All dropout rates are set to 0.5. 
\end{itemize}
Five-fold cross-validation (CV) is performed throughout all experiments. Each fold is a combination of approximately 1/5 tracks of each dataset. Every model is trained on four folds and cross validated on the remaining fold, resulting in a total number of five training/cross-validation scores, the average of which will be the final score to be reported.

We provide all implementation details including the training and cross-validation scripts online \footnote{\url{https://github.com/tangkk/tangkk-mirex-ace}}, so that interested readers can repeat the experiments when they have access to the raw audio datasets.

\section{Results and Discussions} \label{sec:res}

Throughout this section, we use the MIREX ACE standard evaluation metric, weighted chord symbol recall (\textit{WCSR}), to report system performances, where the ``chord symbol recall'' (\textit{CSR}) is defined as follows:
\begin{equation}
\mathit{CSR = {|S\cap S^*|\over|S^*|},}
\label{eq:csr}
\end{equation}
where $S$ and $S^*$ represents the automatic estimated segments, and ground truth annotated segments, respectively, and the intersection of $S$ and $S^*$ are the parts where they overlap and have equal chord annotations. \textit{WCSR} is the weighted average of all tracks' \textit{CSR}s by the lengths of these tracks:
\begin{equation}
\mathit{WCSR = {\sum{Length(Track_i)*CSR_i} \over \sum{Length(Track_i)}}},
\label{eq:wcsr}
\end{equation}
where the subscript $i$ denotes the track number. Unless otherwise specified, we report all \textit{WCSR} scores under the \textit{SeventhsBass} evaluation, where a correct classification does {\it not} involve any chord mapping scheme beyond the \textit{SeventhsBass} vocabulary \cite{pauwels2013evaluating}. We use the MusOOEvaluator tool \footnote{\url{https://github.com/jpauwels/MusOOEvaluator}} to generate these scores from all the ground truth and predicted chord sequences.

The \textit{WCSR} is upper-bounded by the \textit{segmentation quality}, which is computed as {\it directional Hamming distance} (DHD). The DHD from $S^*$ to $S$ is:
\begin{equation}
h(S^*||S) = \sum_{i=1}^{N_{S^*}}(|S_i|-\max_j|S_i^* \cap S_j|),
\end{equation}
where subscription $i$ indicates the $i^{th}$ segment. Note that the distance is not commutable, which means $h(S^*||S)$ and $h(S||S^*)$ represent two different distances. Conventionally, $h(S^*||S)$ measures {\it under-segmentation} and $h(S||S^*)$ measures {\it over-segmentation}. In either case, a good segmentation is indicated by a small value. When reported as scores, they are usually normalized by the lengths of the tracks, and minus by 1, in order to make it equal to the $[0,1]$ range of the \textit{WCSR} score \cite{harte2010towards}:
\begin{equation}
h(S^*,S)=1-{1\over T}\max\{h(S^*||S), h(S||S^*)\} \in [0,1].
\end{equation}
Note that we do not report the segmentation score for every system, because they all share the same GMM-HMM process described in Section~\ref{sec:sg}. This process has a segmentation score of 83\% on the JKURB dataset.

In the following discussion, we will analyze the experiment results from a bias-variance perspective \cite{geman1992neural}. Assuming that a model is trained for multiple times over different samples of the same population, with other settings remain unchanged, the model's prediction for a given input will become a random variable. The model's prediction error can thus be expressed as \cite{friedman2001elements}:
\begin{equation}
Prediction\ Error = Bias^2 + Variance + Irreducible\ Error
\end{equation}
Concretely, a model's bias is defined as the expected difference between its prediction and the ground truth. It measures how much a model's predictions are consistently deviating from the true value, and it indicates whether a model contains fundamentally incorrect assumptions. A model's variance, on the other hand, measures the variance (i.e. the statistical variance, which equals the square of the standard deviation) of the model's prediction. It indicates how much a model's predictions will vary across its different realizations with different training samples, or equivalently, how much inconsistencies are there within the predictions. Finally, the irreducible error term could be seen as collection of everything that is not bias or variance, such as the noise or inconsistencies in the data itself.

Bias and variance is highly correlated with over-fitting and under-fitting. A high bias model tends to under-fit the data, while a high variance model tends to over-fit the data. A model that neither over-fits nor under-fits, will have low bias and low variance. The amount of bias-variance can be approximated from a model's training and validation (or cross-validation) score:
\begin{itemize}
	\item A model with high bias (under-fitting) has similar training and validation scores, but none of them is high.
	\item A model with high variance (over-fitting) has a high training score and a low validation score.
\end{itemize}
In other words, a model's bias can be approximated as the value of its training or validation error if the two errors are close to each other, and a model's variance can be approximated as the difference between its training and validation errors. Note that the irreducible error could either appear as bias or variance.

In the following, we will examine how each design choice in Table~\ref{tab:varexplore} actually affects the systems' biases and variances. Moreover, we also compare among different types of deep neural nets and a baseline model, and see which one performs the best in the LVACE task.

\subsection{Network configurations} \label{sec:p2}

\begin{figure}[h!]
	\centering
	\includegraphics[width=0.6\columnwidth]{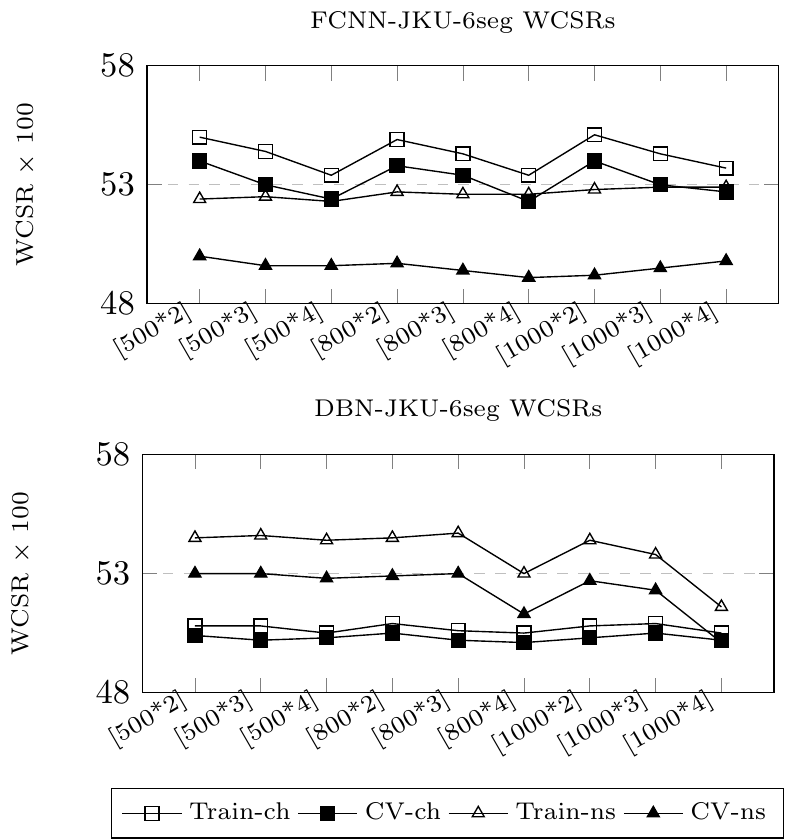}
	\caption{Exploring the effect of different network configurations. All models are trained with JKU-6seg-ch.}
	\label{fig:configs}
\end{figure}

Figure~\ref{fig:configs} shows the \textit{WCSR}s of a set of JKU-6seg models with different neural nets, network configurations and input features.
\Hsection{-ch models}
The FCNN has local maximal validation scores when the network has two layers, and it performs worse as  network becomes deeper. The DBN's validation scores are stabilized around 50. In this group of experiments, the training and validation scores are close to each other.

\Hsection{-ns models}
The FCNN's validation scores are all focused around 50. As for the DBN, there is a trend of performance downgrade along the depth dimension. In both cases the differences between training and validation scores are very small.

\Hsection{Remarks}
Firstly, all the FCNN-ch models outperform the FCNN-ns models. This could be largely due to the prior imposed by the chromagram feature. It embeds the knowledge about ``pitch classes'', and it is originally designed for chord recognition tasks. On the other hand, because the notegram is several feature transformations away from the chromagram, it contains no such prior information. This could explain why the FCNN-ch models perform worse as they become deeper. Every extra layer will tend to weaken the prior at the input, and at the same time, these layers try to learn some other regularities that map the chromagrams to the chord labels. The results show that, unfortunately, the deeper networks are unable to learn more useful regularities than the prior knowledge already contained in the chromagram.

Secondly, all the DBN-ns models outperform the DBN-ch models (except for the one at [1000*4]). Note that the only difference between the DBN and the FCNN is the generative pre-training process, which in effect is a strong regularization process that prevents over-fitting. This is sometimes equivalent to increasing the model's bias or decreasing the model's variance. As shown in Figure~\ref{fig:configs}, since the variances of the FCNN-ch models are already small, the DBN-ch models perform worse than the FCNN-ch models due to the higher biases. However, since the FCNN-ns models have high variances, the DBN-ns models perform better than the FCNN-ns because of the lower variances.

Thirdly, the performance downgrade of the DBN-ns models starting from [800*3] and [1000*2] could be caused by the well-known gradient vanishing problem in deep networks (with sigmoid activations). This could be verified by monitoring the weight updating process. When the gradient vanishing happens, the average amount of weight updates closer to the input will be much less than those closer to the output, resulting in more errors in the earlier layers, which will be aggregated through the feedforward path to the output.

\subsection{Segment tiling} \label{sec:p3}

\begin{figure}[h!]
	\centering
	\includegraphics[width=0.6\columnwidth]{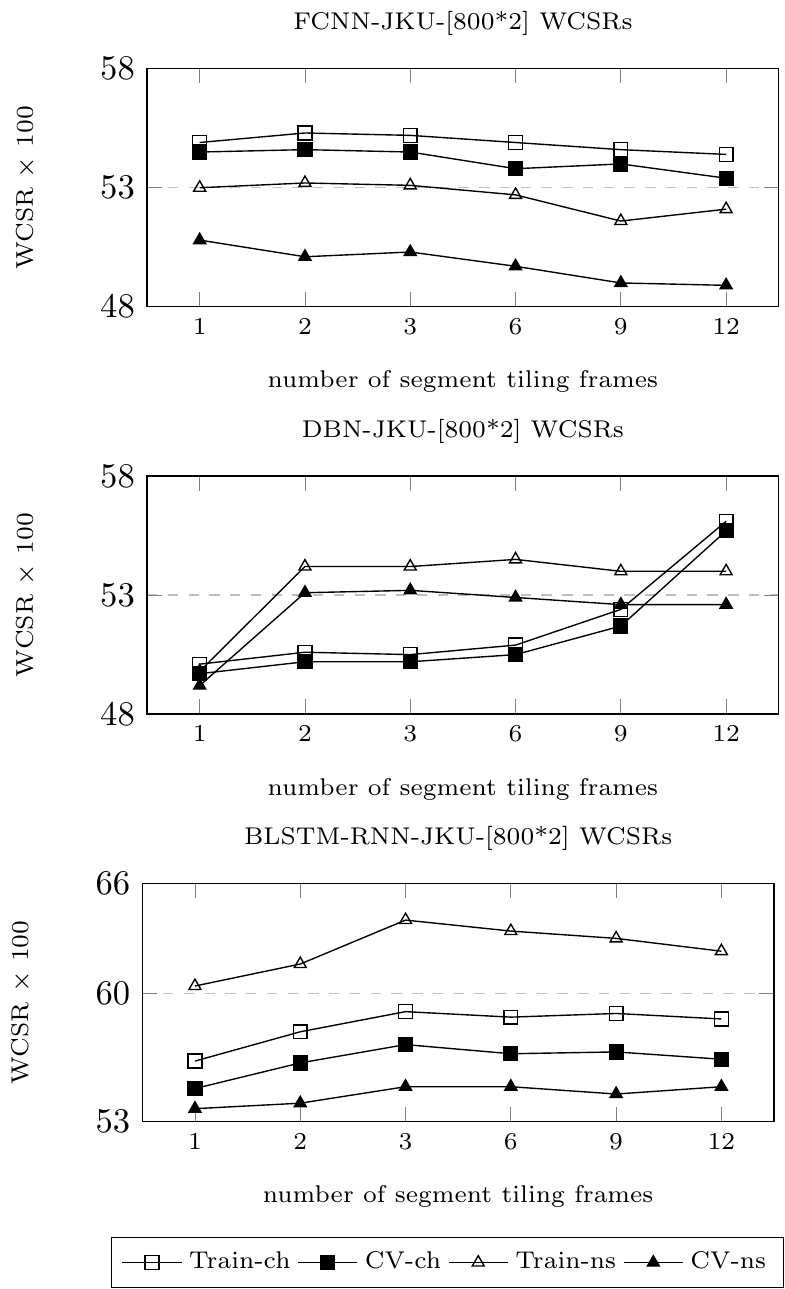}
	\caption{Exploring the effect of segment tiling. All models are trained with JKU-ch-[800*2]}
	\label{fig:nseg}
\end{figure}

Figure~\ref{fig:nseg} shows the \textit{WCSR}s of a set of JKU-[800*2] models with different neural nets, segment tiling schemes ($N$seg) and input features. Note that [800*2] in BLSTM-RNN means there are a forward and a backward hidden layers, each having 800 LSTM units.

\Hsection{-ch models}
The FCNN tends to perform worse with larger $N$; The DBN tends to perform better with larger $N$; The BLSTM-RNN grows gently when $N$ is less than 3, and remains relatively constant thereafter. They all have very small variances.

\Hsection{-ns models}
Still the FCNN tends to perform worse with larger $N$; Both the DBN and the BLSTM-RNN have the worst performances when $N$ is 1, and they have higher and stable performances when $N$ is greater than 1.

\Hsection{Remarks}
With a large $N$seg, a model becomes more complex because of a less blurry input, so that one could expect either less bias, or more variance. In the FCNN models we could observe a tendency of slight variance increasing. This tendency has possibly offset the trend of bias decreasing, which we could not see from Figure~\ref{fig:nseg}.

For the DBN, as discussed in Section~\ref{sec:p2}, the generative pre-training processes could reduce the variances or increase the biases. This is clearly reflected in Figure~\ref{fig:nseg} if we compare the DBN's \textit{WCSR}s with the FCNN's. On one hand, this could explain why the DBN-ch models have an increasing trend of performances (as well as a decreasing trend of biases) with a larger $N$, because the DBN-ch model has a much higher bias than the FCNN-ch one when $N$ equals 1. On the other hand, it could also explain why there is a performance boost from $N=1$ to $N=2$ in the DBN-ns models, and that the DBN-ns models have consistently lower variances and higher performances than the FCNN-ns models.

For the BLSTM-RNN models, the training and CV curves are much more spread out than those of the FCNN and the DBN. On one hand, the RNN imposes a weight sharing mechanism across the segment tiling frames. This has an effect of regularization by limiting the number of parameters connecting the input layer to the hidden layer, thus limiting the network's ability to recognize arbitrary dependencies across frames. On the other hand, the RNN also introduces a set of recurrent weights that connect each frame to its next frame. This makes the network more flexible in capturing sequential dependencies between frames. This explains why the training and CV curves are so separated. Particularly, the average training scores of the RNN models are much higher than those of the DBN's and the FCNN's, because the RNN is essentially biased towards problems with sequential natures, and the ACE is one of these problems. Still, we have relatively low CV scores in these RNN models, which lead to high variances. As we will see in the following subsection, this can be remedied by more training data.

\subsection{Amount of training data}\label{sec:p5}
\begin{figure}[h!]
	\centering
	\includegraphics[width=0.6\columnwidth]{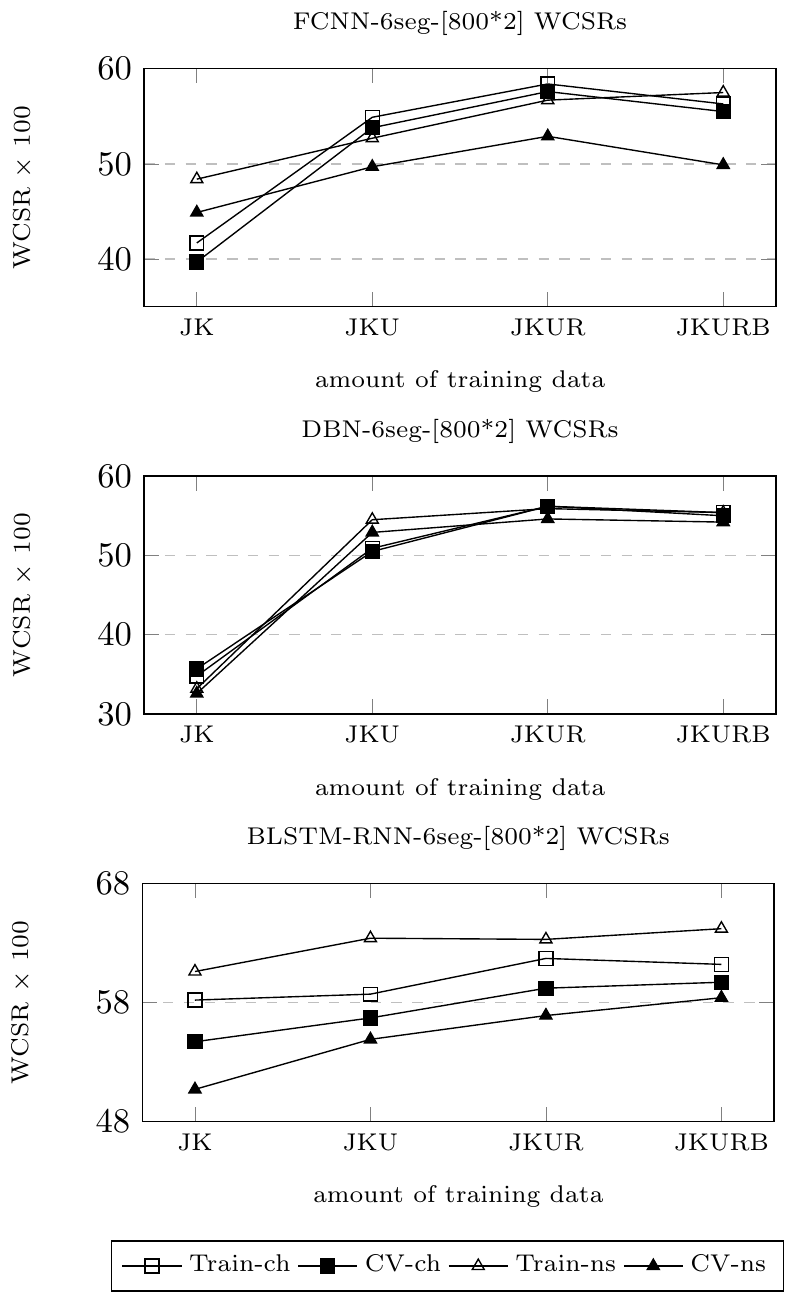}
	\caption{Exploring the different training data size. All models are trained with 6seg-ch-[800*2].}
	\label{fig:data}
\end{figure}

Figure~\ref{fig:data} shows the \textit{WCSR}s of a set of 6seg-[800*2] models with different neural nets, training data sizes and input features.

\Hsection{For both -ch and -ns models}
In all three plots, there are clear trends that increasing the amount of data boosts the models' performances. While the variances of the FCNN and DBN models remain being small, the variances of the BLSTM-RNN models tend to decrease with the increase of data. Interestingly for the FCNN, the increase of data from JKUR to JKURB leads to larger variances and worse CV scores. 

\Hsection{Remarks}
Similar to Figure~\ref{fig:nseg}, the FCNN's and DBN's training and CV curves as shown in Figure~\ref{fig:data} are still very close to each other. It seems that as the amount of data increases, their models have saturated at some point and there is little room for further improvement. On the contrary, the BLSTM-RNN’s training and CV curves are much wider apart, and the models tend to generalize better as the data size grows. The results in Figure~\ref{fig:data} seem to suggest that we have almost touched a performance ``ceiling'' of the BLSTM-RNN-ch models, but we have yet to reach that of the BLSTM-RNN-ns models.

\subsection{Input feature}\label{sec:p6}
From Figure~\ref{fig:configs} to~\ref{fig:data}, we see that the training and CV curves of the chromagram models are on average much closer than those of the notegram models. This suggests that the prior knowledge contained in the chromagram feature actually introduces bias in the model. On one hand, this may lead to better models if the amount training data is limited (this discussion is not valid for the DBN because of the generative pre-training process). On the other hand, this can also limit the models' improvement when we have sufficient amount of training data. For example in Figure~\ref{fig:data}, we see a potential trend that if more data is added to the model, the BLSTM-RNN-ns model will eventually outperform the BLSTM-RNN-ch model, because the BLSTM-RNN-ns has a higher ceiling (the training score) than the BLSTM-RNN-ch.

\subsection{Balanced performance}\label{sec:p7}
The above discussions are focused on the overall \textit{WCSR}s of different models. Here we are going to examine the models' performances on specific chords. Note that in our datasets (which we believe are good representatives of pop and rock music in general), the chord distributions are highly skewed (as shown in Table~\ref{tab:chorddist}), where the $maj$ and $min$ triads make up almost 70\% of the whole sample population, the $maj7$, $min7$ and $7$ chords constitute more than 20\%, and the portion of other chords are less than 10\%. In the following discussion, we refer to ``common chords'' as the $maj$ and $min$ chords, ``uncommon chords'' as the sevenths chords, including the $maj7$, $min7$ and $7$ chords, and ``long-tail chords'' as all the other chords in the \textit{SeventhsBass} vocabulary. Moreover, we use ``chord'' and ``chord type'' interchangeably to refer to a certain type of chords. We report system performance on chords using the per chord \textit{WCSR}:
\begin{equation}
\mathit{WCSR_{C} = {\sum{Length(C_i)*CSR_i} \over \sum{Length(C_i)}}},
\label{eq:wcsrchord}
\end{equation}
where the subscript $i$ denotes the $i^{th}$ instance of the chord $C$ within the data set.

\begin{figure}[h!]
	\centering
	\includegraphics[width=1\columnwidth]{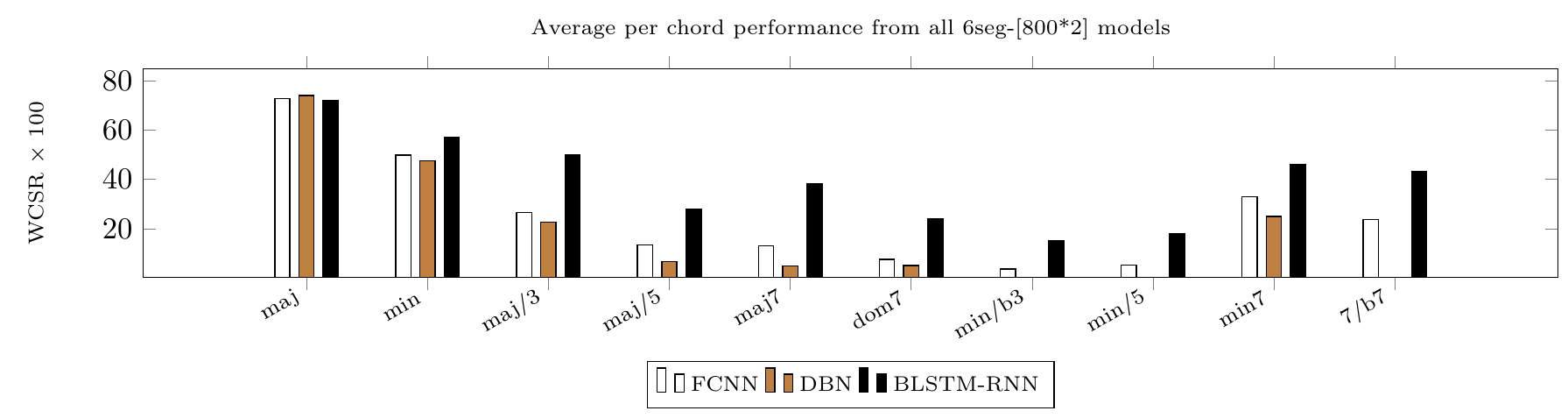}
	\caption{Performance on different chords in different neural nets. All models are trained with 6seg-[800*2].}
	\label{fig:skewed}
\end{figure}

Figure~\ref{fig:skewed} shows how different deep neural net models perform on different chords. It is surprising to see that the FCNN and the DBN outperform the BLSTM-RNN only in the \textit{maj} chord category, while the BLSTM-RNN outscores the other two by large margins in most long-tail chords and uncommon chords categories.

\begin{figure}[h!]
	\centering
	\includegraphics[width=0.4\columnwidth]{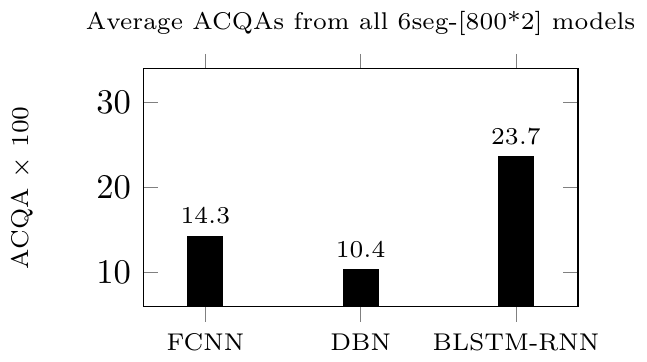}
	\caption{Average ACQAs of \textit{SeventhsBass} Vocabulary. All models are trained with 6seg-[800*2].}
	\label{fig:acqa}
\end{figure}

Furthermore, we examine the versatilities of different deep neural net models. We measure them using the ``Average Chord Quality Accuracy''(\textit{ACQA}) \cite{cho2014improved}, which averages the \textit{WCSR}s of all chords with equal weights:
\begin{equation}
\mathit{ACQA = {{\sum{WCSR_{C}}} \over {\#\,of\,chords}}}.
\label{eq:acqa}
\end{equation}
Models that over-fit a few chord types tend to give lower \textit{ACQA}s, while those well-balanced ones will have higher \textit{ACQA}s. As shown in Figure~\ref{fig:acqa}, the average \textit{ACQA} of the BLSTM-RNN models outscores the average \textit{ACQA}s of the other two types of models by around 10 points.

\begin{figure}[h!]
	\centering
	\includegraphics[width=0.8\columnwidth]{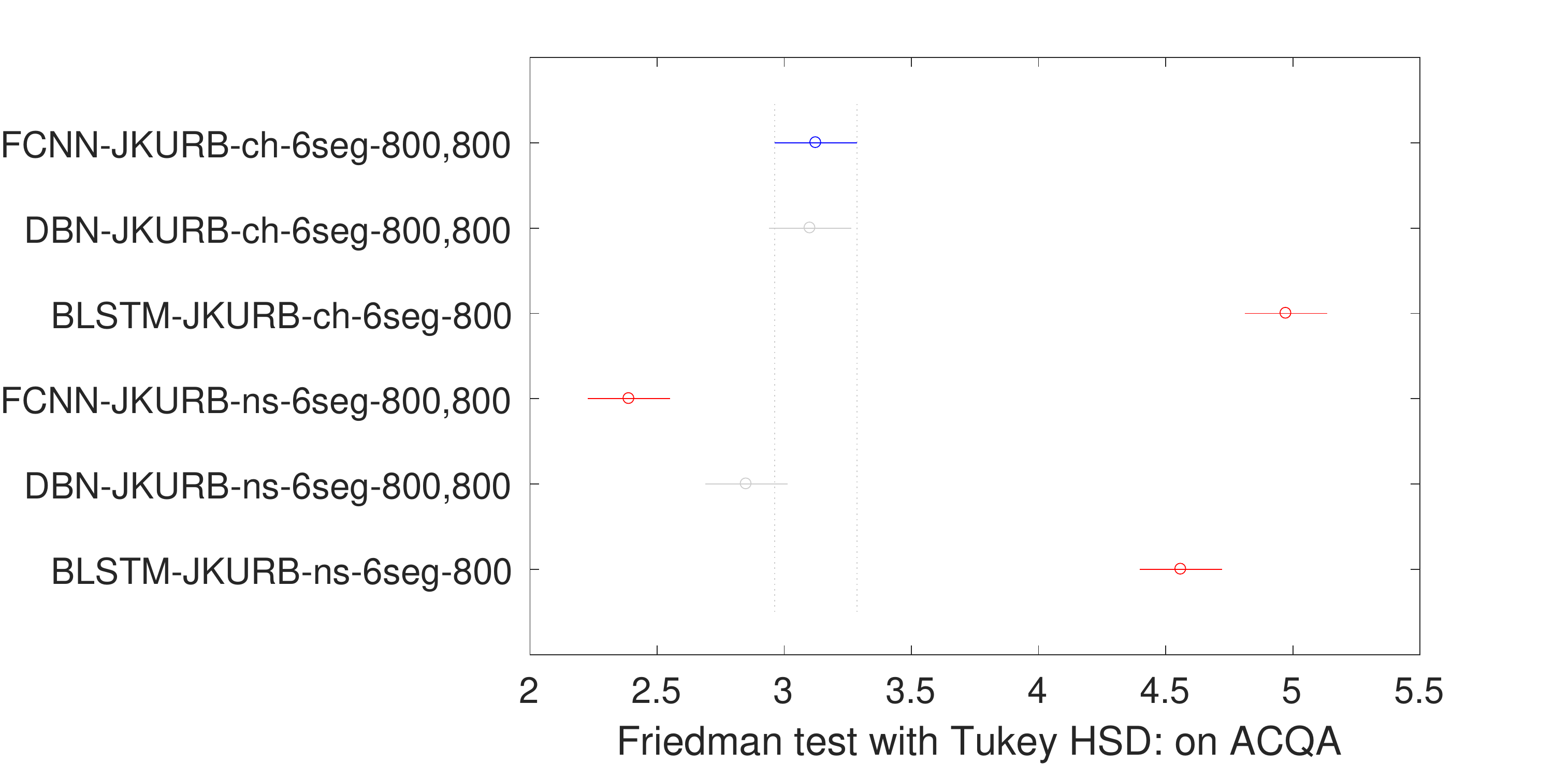}
	\caption{Friedman test with Tukey HSD: ACQAs of different system variants}
	\label{fig:friedmanacqa}
\end{figure}

We perform a Friedman test \cite{friedman1937use} on the track-wise \textit{ACQA} results. After that we use the Tukey HSD (honest significant difference) \cite{tukey1949comparing} to perform a multiple comparison test on the Friedman test's statistics with a significance level of 0.05. As shown in Figure~\ref{fig:friedmanacqa}, both BLSTM-JKURB-ch-6seg-800 and BLSTM-JKURB-ns-6seg-800 are significantly better (no overlap of confidence intervals) than the other systems, and BLSTM-JKURB-ch-6seg-800 is significantly better than BLSTM-JKURB-ns-6seg-800 as well. This concludes that the BLSTM-RNN models are significantly better than the FCNN and the DBN models in terms of \textit{ACQA}s, or balanced performances.

Now we have concrete evidence that the BLSTM-RNN is a better neural network in solving the LVACE problem than the other two models. It is reasonable to think that the BLSTM-RNN regards its input as a {\it sequence of frames}, while fully-connected networks (in this context the FCNN and DBN) regard their inputs as \textit{flat vectors}. Therefore, while the BLSTM-RNN tries to look for regularities within \textit{each pair of consecutive frames} along the time direction, the FCNN or DBN would search for regularities within \textit{every point of the flat vector} as if they are not time related at all. Another perspective is that the BLSTM-RNN has $1/N$ times the weights as much as those of the FCNN's and the DBN's between the input layer and the first hidden layer. Thus the weight sharing over multiple frames prevents the BLSTM-RNN from over-fitting, and allows the model to process higher resolution inputs without an increase in parameters.

In some cases, the fully-connected network is more efficient given that the input feature has already encoded certain prior information about music (e.g. chromagram contains the information about pitch classes). Nevertheless, it overlooks the ``sequential order of frames'', which probably causes the over-fitting of root position chords and the under-fitting of chord inversions.

\subsection{Baseline comparison} \label{sec:p9}
Finally, we compare our LVACE framework with the Chordino. It should be emphasized that Chordino is the only suitable baseline because: (1) Our framework resembles Chordino in terms of the segmentation and feature extraction processes; (2) Chordino is the only other system that supports seventh chords and chord inversions.
\begin{figure}[h!]
	\centering
	\includegraphics[width=0.6\columnwidth]{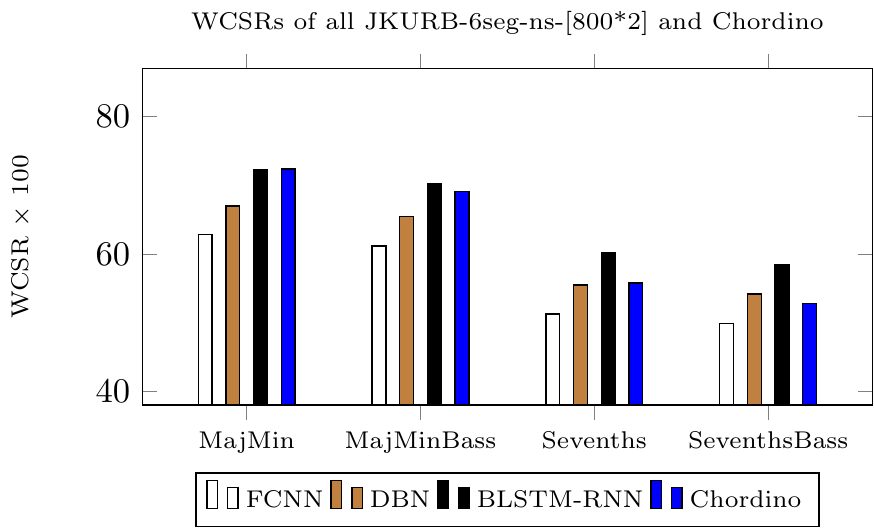}
	\caption{Performance comparison between system representatives and Chordino on WCSRs. All models are trained with JKURB-6seg-ns-[800*2].}
	\label{fig:compchordino}
\end{figure}

We choose one representative for each type of deep neural net, all trained and cross-validated with JKURB-6seg-ns-[800*2], and compare them with the Chordino using the standard MIREX ACE categories as described in Section~\ref{sec:aceeval}. As shown in Figure~\ref{fig:compchordino}, the representative of BLSTM-RNN outperforms the Chordino by large margin in \textit{Sevenths} and \textit{SeventhsBass}, and it scores fairly close to the Chordino in \textit{MajMin} and \textit{MajMinBass}. The other two representatives are not performing as good as the Chordino in most categories.

\begin{figure}[h!]
	\centering
	\includegraphics[width=0.8\columnwidth]{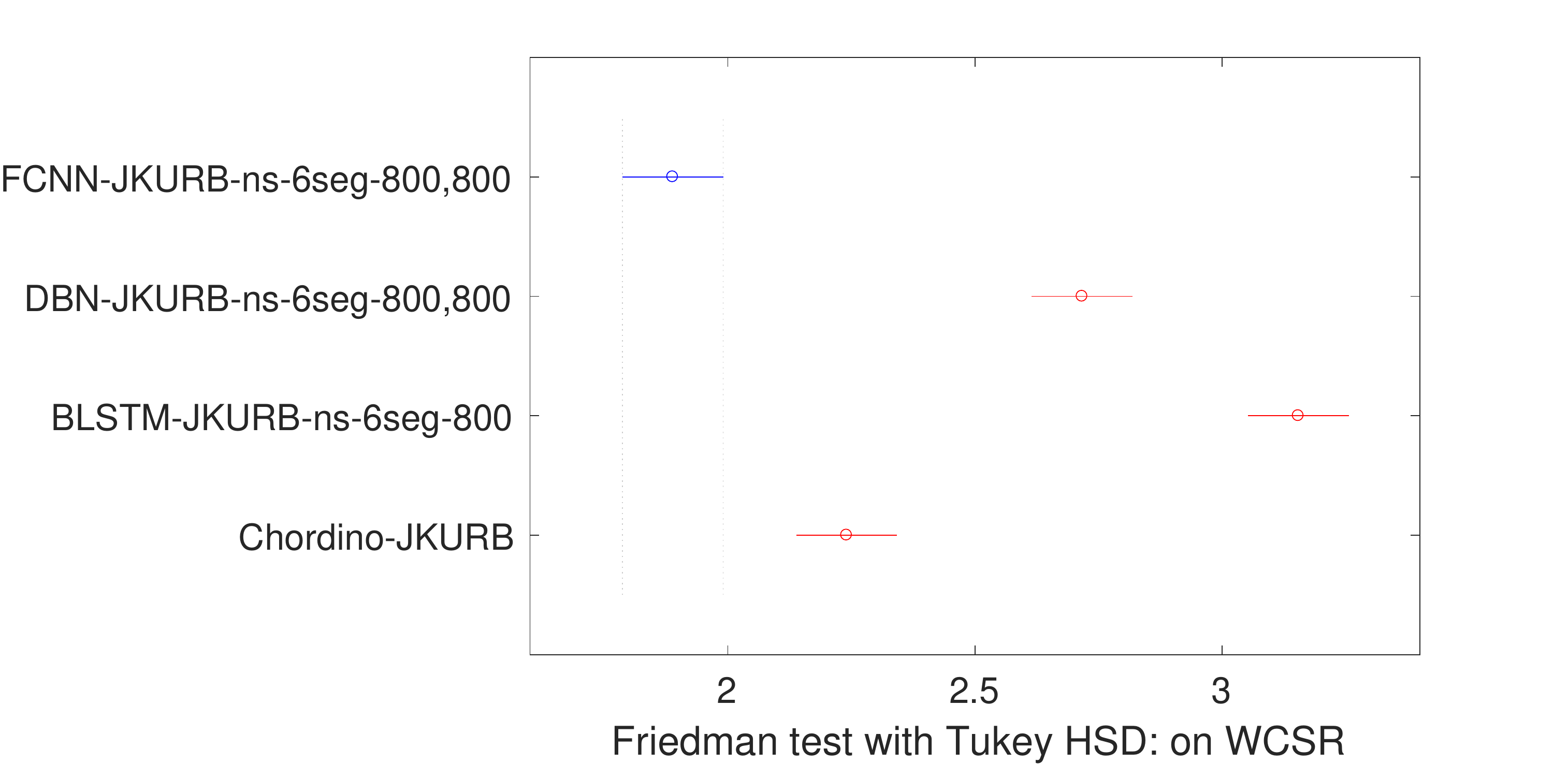}
	\caption{Friedman test with Tukey HSD: WCSRs compared with the baseline}
	\label{fig:friedmanwcsrbl}
\end{figure}

We perform a Friedman test on the track-wise \textit{SeventhsBass} \textit{WCSR} results. After that we use the Tukey HSD to perform a multiple comparison test on the Friedman test's statistics with a significance level of 0.05. As shown in Figure~\ref{fig:friedmanwcsrbl}, BLSTM-JKURB-ns-6seg-[800*2] is significantly better than the other systems as well as the Chordino.

\begin{figure}[h!]
	\centering
	\includegraphics[width=0.4\columnwidth]{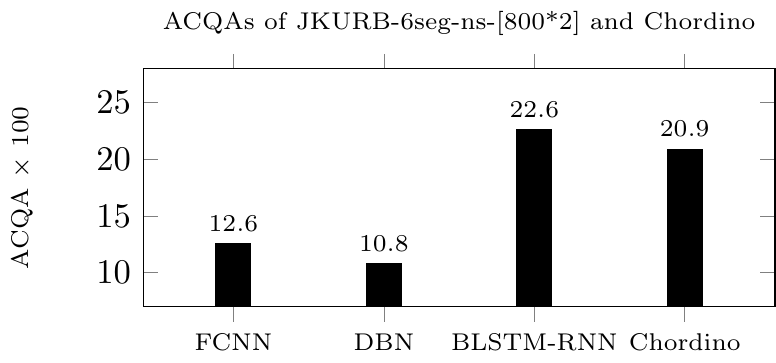}
	\caption{Performance comparison between system representatives and Chordino on ACQAs. All models are trained with JKURB-6seg-ns-[800*2].}
	\label{fig:acqachordino}
\end{figure}

\begin{figure}[h!]
	\centering
	\includegraphics[width=0.8\columnwidth]{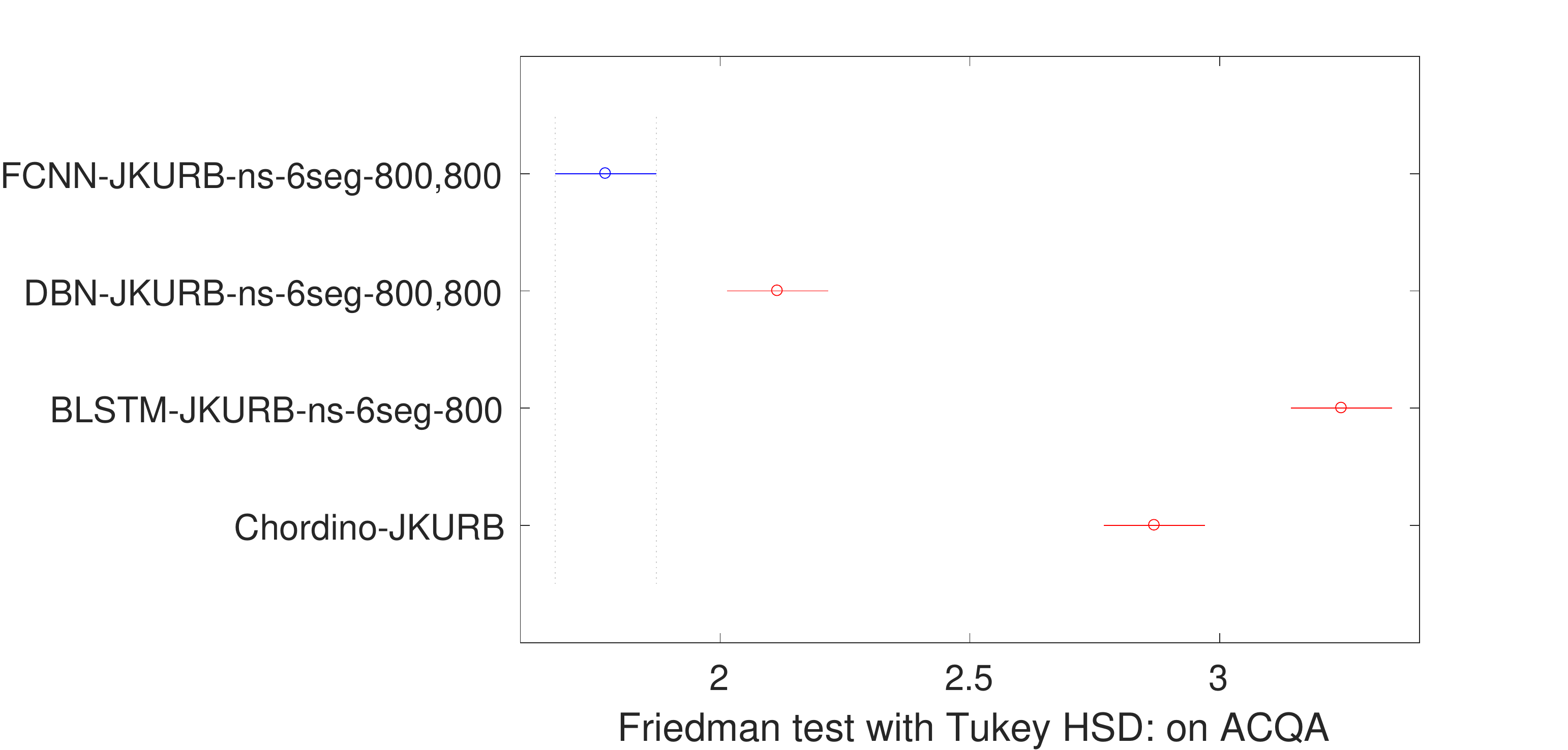}
	\caption{Friedman test with Tukey HSD: ACQAs compared with the baseline}
	\label{fig:friedmanacqabl}
\end{figure}

In terms of \textit{ACQA}, as shown in Figure~\ref{fig:acqachordino}, Chordino outperforms both the FCNN's and DBN's representatives, but the most balanced system is the BLSTM-RNN's representative. We again perform a Friedman test with Tukey HSD ($p=0.05$, using the track-wise results) to test whether the differences in \textit{ACQA}s are significant. As shown in Figure~\ref{fig:friedmanacqabl}, the BLSTM-JKURB-ns-6seg-[800*2] system is again significantly better than the other systems as well as the Chordino.

\section{Conclusions} \label{sec:concln}

In this paper we present an in-depth discussion of a hybrid ``GMM-HMM + deep neural net''  LVACE approach. Preluded with an argument for the necessity of recognizing chord inversions in practical ACE systems, our work is motivated by a current research gap in ACE, which is the overlooking of large vocabulary and chord inversions. This is the rationale behind the \textit{SeventhsBass} LVACE implementation. We then put forward the LVACE system framework, which has handcrafted feature extraction and segmentation processes, and uses deep neural nets to classify chords from the features. We conduct several groups of experiments on different system variants of the LVACE framework, from which we report the following major findings:
\begin{itemize}
	\item The chromagram feature contains prior knowledge about musical pitch class that increases the bias and limits the potential improvement of the models.
	\item The BLSTM-RNN can learn regularities from the notegram feature that potentially outperforms the chromagram feature.
	\item The BLSTM-RNN's representative system (with all available training data) has significantly better \textit{WCSR} and \textit{ACQA} than the FCNN's one, the DBN's one, and the Chordino.
\end{itemize}

Despite the best system variant significantly outperforms the baseline system, all training and CV scores presented in this paper are still far less than 100\%. This indicates either there is large bias in the LVACE framework itself, or there is irreducible error in the underlying data. We speculate three potential causes as explained in the following.

Firstly, the performance of the proposed framework is upper-bounded by the segmentation performance of the GMM-HMM process introduced in Section~\ref{sec:sg}, and the performance of this process on the JKURB set is 83\%.

Secondly, the segment tiling process introduces bias to the system, since it assumes a chord can be correctly recognized after we tile its original features into several frames of averaged features. This process could help prevent over-fitting by regularizing the degree of freedom of the input, but at times it scarifies important information conveyed in the original variable-length features.

The above two points set a hard performance limit of the proposed LVACE framework: unless the chord segmentation technique is perfect and the segment tiling process is completely excluded, one could not expect a system with very low bias.

Thirdly, there is non-negligible amount of noise in the ground truth annotations themselves. Inevitably, due to differences in musical training, human annotators sometimes disagree, particularly on long-tail chords \cite{humphreyfour}. This results in a glass ceiling for LVACE: unless there are more data for uncommon and long-tail chords and they are more consistently labeled, all efforts for improving LVACE will be hindered by the lack of skewed class training and data consistency. 

In a very strict sense, there is not any ``gold standard'' if human annotators themselves might disagree with each other. But in a loose sense, there could be a ``gold standard'' if:
\begin{itemize}
	\item all annotations are done by only one annotator, or
	\item all annotations are done by multiple annotators (much more than two).
\end{itemize}
In the former case, the only annotator ``dictates'' a local ``gold standard'', so that whenever a machine tries to learn from the data, it actually targets at this annotator's ``style''. In the latter case, multiple annotators decide a ``gold standard'' in a way such as majority vote or data fusion \cite{koopsintegration,klein2004sensor}, so that a trained model actually aims at the optimal ``style'' that minimizes the objections among these annotators. Therefore, although the ``gold standard'' is indeed an important issue, we still have to design a system that ``learns well''.

We believe that the next step of LVACE research should focus more on improving the recognition accuracies on uncommon and long-tail chords. That is, instead of considering the overall \textit{WCSR} of a large vocabulary, attention should also be given to the balanced metric, such as \textit{ACQA}. Although we have pointed out that the BLSTM-RNN is very promising in handling large vocabulary with inversions, we have yet to explored possible ways to train the network under such ``imbalanced class population '' scenario \cite{chawla2004editorial}. More importantly, we should spend greater efforts on data collection, particularly of long-tail chords, and at the same time ensure the data integrity and consistency, in the future development of LVACE.

\begin{landscape}
	\vspace*{\fill}
	\begin{table*}[h!]
		\tiny
		\label{tab:chorddist}
		\begin{tabular}{|c|c|c|c|c|c|c|c|c|c|c|c|c|c|c|c|c|c|c|c|}\hline
			Dataset & Tracks & M/5 & M/3 & M & M7/5 & M7/3 & M7/7 & M7 & 7/5 & 7/3 & 7/b7 & 7 & m/5 & m/b3 & m & m7/5 & m7/b3 & m7/b7 & m7\\ \hline
			C & 20 & 3.2 & 4.5 & 38.4 & 0.2 & 0.2 & 0.0 & 10.2 & 0.0 & 0.3 & 1.1 & 9.3 & 2.7 & 0.0 & 20.3 & 0.5 & 0.0 & 0.1 & 9.1\\ \hline
			J & 29 & 4.1 & 8.1 & 32.3 & 1.2 & 0.2 & 0.1 & 6.9 & 0.4 & 1.5 & 2.3 & 5.0 & 0.7 & 1.3 & 15.1 & 0.7 & 0.0 & 0.3 & 19.8\\ \hline
			K & 26 & 4.5 & 3.9 & 52.0 & 0.0 & 0.3 & 0.2 & 5.1 & 0.3 & 0.2 & 0.5 & 6.2 & 0.1 & 0.3 & 14.9 & 0.2 & 0.0 & 0.8 & 10.4\\ \hline
			U & 191 & 2.3 & 3.9 & 54.7 & 0.0 & 0.1 & 0.1 & 3.2 & 0.1 & 0.3 & 0.3 & 8.3 & 0.4 & 0.4 & 15.1 & 0.0 & 0.1 & 0.4 & 10.2\\ \hline
			R & 100 & 2.5 & 4.6 & 42.8 & 0.3 & 0.1 & 0.1 & 8.9 & 0.0 & 0.2 & 0.3 & 7.9 & 0.4 & 0.5 & 15.3 & 0.0 & 0.1 & 0.2 & 15.7\\ \hline
			B & 180 & 2.4 & 1.0 & 66.1 & 0.0 & 0.2 & 0.3 & 0.9 & 0.1 & 0.1 & 0.4 & 8.7 & 0.6 & 0.5 & 15.9 & 0.0 & 0.1 & 0.4 & 2.5\\ \hline
			WA & & 2.6 & 3.3 & 54.3 & 0.1 & 0.1 & 0.2 & 4.0 & 0.1 & 0.3 & 0.5 & 8.1 & 0.6 & 0.5 & 15.6 & 0.1 & 0.1 & 0.4 & 9.1\\ \hline
		\end{tabular}
		\caption{Distribution of chords (M=maj, m=min) in the datasets and their weighted averages (WA) by the number of tracks. (maj and min: 69.9\%; maj7, min7 and 7: 21.3\%; others: 8.8\%)}
	\end{table*}
	\vspace*{\fill}
\end{landscape}

\bibliography{references}

\begin{thebibliography}{57}
\providecommand{\natexlab}[1]{#1}
\providecommand{\url}[1]{\texttt{#1}}
\expandafter\ifx\csname urlstyle\endcsname\relax
  \providecommand{\doi}[1]{doi: #1}\else
  \providecommand{\doi}{doi: \begingroup \urlstyle{rm}\Url}\fi

\bibitem[Bello(2007)]{bello2007audio}
Juan~Pablo Bello.
\newblock Audio-based cover song retrieval using approximate chord sequences:
  Testing shifts, gaps, swaps and beats.
\newblock In \emph{Proceedings of the 8th International Society for Music
  Information Retrieval Conference, ISMIR 2007}, volume~7, pages 239--244,
  2007.

\bibitem[Lee(2006)]{lee2006identifying}
Kyogu Lee.
\newblock Identifying cover songs from audio using harmonic representation.
\newblock \emph{extended abstract, Music Information Retrieval eXchange task,
  Victoria, BC, Canada}, 2006.

\bibitem[Serra et~al.(2010)Serra, G{\'o}mez, and Herrera]{serra2010audio}
Joan Serra, Emilia G{\'o}mez, and Perfecto Herrera.
\newblock Audio cover song identification and similarity: background,
  approaches, evaluation, and beyond.
\newblock pages 307--332, 2010.

\bibitem[Bello and Pickens(2005)]{bello2005robust}
Juan~Pablo Bello and Jeremy Pickens.
\newblock A robust mid-level representation for harmonic content in music
  signals.
\newblock In \emph{Proceedings of the 6th International Society for Music
  Information Retrieval Conference, ISMIR}, volume~5, pages 304--311, 2005.

\bibitem[Cheng et~al.(2008)Cheng, Yang, Lin, Liao, and
  Chen]{cheng2008automatic}
Heng-Tze Cheng, Yi-Hsuan Yang, Yu-Ching Lin, I-Bin Liao, and Homer~H Chen.
\newblock Automatic chord recognition for music classification and retrieval.
\newblock In \emph{IEEE International Conference on Multimedia and Expo}, pages
  1505--1508. IEEE, 2008.

\bibitem[P{\'e}rez-Sancho et~al.(2009)P{\'e}rez-Sancho, Rizo, and
  Inesta]{perez2009genre}
Carlos P{\'e}rez-Sancho, David Rizo, and Jos{\'e}~M Inesta.
\newblock Genre classification using chords and stochastic language models.
\newblock \emph{Connection science}, 21\penalty0 (2-3):\penalty0 145--159,
  2009.

\bibitem[Papadopoulos and Tzanetakis(2012)]{papadopoulos2012modeling}
H{\'e}l{\`e}ne Papadopoulos and George Tzanetakis.
\newblock Modeling chord and key structure with {Markov} logic.
\newblock In \emph{Proceedings of the 13th International Society for Music
  Information Retrieval Conference, ISMIR}, pages 127--132. Citeseer, 2012.

\bibitem[Pauwels and Martens(2010)]{pauwels2010integrating}
Johan Pauwels and Jean-Pierre Martens.
\newblock Integrating musicological knowledge into a probabilistic framework
  for chord and key extraction.
\newblock In \emph{Audio Engineering Society Convention 128}. Audio Engineering
  Society, 2010.

\bibitem[Papadopoulos and Peeters(2008)]{papadopoulos2008simultaneous}
H{\'e}lene Papadopoulos and Geoffroy Peeters.
\newblock Simultaneous estimation of chord progression and downbeats from an
  audio file.
\newblock In \emph{IEEE International Conference on Acoustics, Speech and
  Signal Processing. ICASSP}, pages 121--124. IEEE, 2008.

\bibitem[Mauch and Dixon(2010{\natexlab{a}})]{mauch2010simultaneous}
Matthias Mauch and Simon Dixon.
\newblock Simultaneous estimation of chords and musical context from audio.
\newblock \emph{IEEE Transactions on Audio, Speech, and Language Processing},
  18\penalty0 (6):\penalty0 1280--1289, 2010{\natexlab{a}}.

\bibitem[Huang et~al.(2001)Huang, Acero, Hon, and Foreword
  By-Reddy]{huang2001spoken}
Xuedong Huang, Alex Acero, Hsiao-Wuen Hon, and Raj Foreword By-Reddy.
\newblock Spoken language processing: A guide to theory, algorithm, and system
  development.
\newblock 1, 2001.

\bibitem[Deng(2006)]{deng2006dynamic}
Li~Deng.
\newblock Dynamic speech models: theory, algorithms, and applications.
\newblock \emph{Synthesis Lectures on Speech and Audio Processing}, 2\penalty0
  (1):\penalty0 1--118, 2006.

\bibitem[Deng and Yu(2014)]{deng2014deep}
Li~Deng and Dong Yu.
\newblock Deep learning: methods and applications.
\newblock \emph{Foundations and Trends in Signal Processing}, 7\penalty0
  (3--4):\penalty0 197--387, 2014.

\bibitem[Yu and Deng(2011)]{yu2011deep}
Dong Yu and Li~Deng.
\newblock Deep learning and its applications to signal and information
  processing [exploratory dsp].
\newblock \emph{Signal Processing Magazine, IEEE}, 28\penalty0 (1):\penalty0
  145--154, 2011.

\bibitem[Sheh and Ellis(2003)]{sheh2003chord}
Alexander Sheh and Daniel~PW Ellis.
\newblock Chord segmentation and recognition using {EM}-trained hidden {Markov}
  models.
\newblock In \emph{Proceedings of the 4th International Society for Music
  Information Retrieval Conference, ISMIR}, pages 185--191. International
  Symposium on Music Information Retrieval, 2003.

\bibitem[Fujishima(1999)]{fujishima1999realtime}
Takuya Fujishima.
\newblock Realtime chord recognition of musical sound: A system using common
  lisp music.
\newblock In \emph{Proceedings of the 25th International Computer Music
  Conference}, volume 1999, pages 464--467, 1999.

\bibitem[Wakefield(1999)]{wakefield1999mathematical}
Gregory~H Wakefield.
\newblock Mathematical representation of joint time-chroma distributions.
\newblock In \emph{SPIE's International Symposium on Optical Science,
  Engineering, and Instrumentation}, pages 637--645. International Society for
  Optics and Photonics, 1999.

\bibitem[Weller et~al.(2009)Weller, Ellis, and Jebara]{weller2009structured}
Adrian Weller, Daniel Ellis, and Tony Jebara.
\newblock Structured prediction models for chord transcription of music audio.
\newblock In \emph{International Conference on Machine Learning and
  Applications, ICMLA}, pages 590--595. IEEE, 2009.

\bibitem[Khadkevich and Omologo(2011)]{khadkevich2011time}
Maksim Khadkevich and Maurizio Omologo.
\newblock Time-frequency reassigned features for automatic chord recognition.
\newblock In \emph{IEEE International Conference on Acoustics, Speech and
  Signal Processing (ICASSP)}, pages 181--184. IEEE, 2011.

\bibitem[Ni et~al.(2012)Ni, McVicar, Santos-Rodriguez, and De~Bie]{ni2012end}
Yizhao Ni, Matt McVicar, Raul Santos-Rodriguez, and Tijl De~Bie.
\newblock An end-to-end machine learning system for harmonic analysis of music.
\newblock \emph{IEEE Transactions on Audio, Speech, and Language Processing},
  20\penalty0 (6):\penalty0 1771--1783, 2012.

\bibitem[Cho and Bello(2013)]{cho2013mirex}
Taemin Cho and Juan~P Bello.
\newblock {MIREX} 2013: Large vocabulary chord recognition system using
  multi-band features and a multi-stream {HMM}.
\newblock \emph{Music Information Retrieval Evaluation {eXchange} ({MIREX})},
  2013.

\bibitem[Burgoyne et~al.(2007)Burgoyne, Pugin, Kereliuk, and
  Fujinaga]{burgoyne2007cross}
John~Ashley Burgoyne, Laurent Pugin, Corey Kereliuk, and Ichiro Fujinaga.
\newblock A cross-validated study of modelling strategies for automatic chord
  recognition in audio.
\newblock In \emph{Proceedings of the 8th International Society for Music
  Information Retrieval Conference, ISMIR}, pages 251--254, 2007.

\bibitem[Mauch and Dixon(2010{\natexlab{b}})]{mauch2010mirex}
Matthias Mauch and Simon Dixon.
\newblock {MIREX} 2010: Chord detection using a dynamic {Bayesian} network.
\newblock \emph{Music Information Retrieval Evaluation Exchange ({MIREX})},
  2010{\natexlab{b}}.

\bibitem[Mauch(2010{\natexlab{a}})]{mauchsimple}
Matthias Mauch.
\newblock Simple chord estimate: Submission to the {MIREX} chord estimation
  task.
\newblock \emph{Music Information Retrieval Evaluation Exchange ({MIREX})},
  2010{\natexlab{a}}.

\bibitem[Humphrey and Bello(2012)]{humphrey2012rethinking}
Eric~J Humphrey and Juan~P Bello.
\newblock Rethinking automatic chord recognition with convolutional neural
  networks.
\newblock In \emph{Proceedings of the 11th International Conference on Machine
  Learning and Applications (ICMLA)}, volume~2, pages 357--362. IEEE, 2012.

\bibitem[Boulanger-Lewandowski et~al.(2013)Boulanger-Lewandowski, Bengio, and
  Vincent]{boulanger2013audio}
Nicolas Boulanger-Lewandowski, Yoshua Bengio, and Pascal Vincent.
\newblock Audio chord recognition with recurrent neural networks.
\newblock In \emph{Proceedings of the 14th International Society for Music
  Information Retrieval Conference, ISMIR}, pages 335--340, 2013.

\bibitem[Sigtia et~al.(2015)Sigtia, Boulanger-Lewandowski, and
  Dixon]{sigtia2015audio}
Siddharth Sigtia, Nicolas Boulanger-Lewandowski, and Simon Dixon.
\newblock Audio chord recognition with a hybrid recurrent neural network.
\newblock In \emph{Proceedings of the 16th International Society for Music
  Information Retrieval Conference (ISMIR)}, 2015.

\bibitem[Zhou and Lerch(2015)]{zhou2015chord}
Xinquan Zhou and Alexander Lerch.
\newblock Chored detection using deep learning.
\newblock In \emph{Proceedings of the 16th International Society for Music
  Information Retrieval Conference, ISMIR}, volume~53, 2015.

\bibitem[Downie(2008)]{downie2008music}
J~Stephen Downie.
\newblock The music information retrieval evaluation exchange (2005-2007): A
  window into music information retrieval research.
\newblock \emph{Acoustical Science and Technology}, 29\penalty0 (4):\penalty0
  247--255, 2008.

\bibitem[Pauwels and Peeters(2013)]{pauwels2013evaluating}
Johan Pauwels and Geoffroy Peeters.
\newblock Evaluating automatically estimated chord sequences.
\newblock In \emph{IEEE International Conference on Acoustics, Speech and
  Signal Processing (ICASSP)}, pages 749--753. IEEE, 2013.

\bibitem[Raffel et~al.(2014)Raffel, McFee, Humphrey, Salamon, Nieto, Liang,
  Ellis, and Raffel]{raffel2014mir_eval}
Colin Raffel, Brian McFee, Eric~J Humphrey, Justin Salamon, Oriol Nieto, Dawen
  Liang, Daniel~PW Ellis, and C~Colin Raffel.
\newblock mir\_eval: A transparent implementation of common mir metrics.
\newblock In \emph{Proceedings of the 15th International Society for Music
  Information Retrieval Conference, ISMIR}. Citeseer, 2014.

\bibitem[Burgoyne et~al.(2014)Burgoyne, de~Haas, and
  Pauwels]{burgoyne2014comparative}
J~Ashley Burgoyne, W~Bas de~Haas, and Johan Pauwels.
\newblock On comparative statistics for labelling tasks: What can we learn from
  {MIREX} {ACE} 2013.
\newblock In \emph{Proceedings of the 15th Conference of the International
  Society for Music Information Retrieval (ISMIR)}, pages 525--530, 2014.

\bibitem[Deng and Kwok(2016)]{deng2016chord}
Junqi Deng and Yu-Kwong Kwok.
\newblock Automatic chord estimation on {SeventhsBass} chord vocabulary using
  deep neural network.
\newblock In \emph{Proceedings of the 41th International Conference on
  Acoustics, Speech and Signal Processing (ICASSP)}, 2016.

\bibitem[Mauch(2010{\natexlab{b}})]{mauch2010automatic}
Matthias Mauch.
\newblock \emph{Automatic chord transcription from audio using computational
  models of musical context}.
\newblock PhD thesis, School of Electronic Engineering and Computer Science
  Queen Mary, University of London, 2010{\natexlab{b}}.

\bibitem[Mauch and Dixon(2010{\natexlab{c}})]{mauch2010approximate}
Matthias Mauch and Simon Dixon.
\newblock Approximate note transcription for the improved identification of
  difficult chords.
\newblock In \emph{Proceedings of the 11th International Society for Music
  Information Retrieval Conference, ISMIR}, pages 135--140, 2010{\natexlab{c}}.

\bibitem[G{\'o}mez(2006)]{gomez2006tonal_b}
Emilia G{\'o}mez.
\newblock Tonal description of music audio signals.
\newblock \emph{Department of Information and Communication Technologies},
  2006.

\bibitem[Lawson and Hanson(1995)]{lawson1995solving}
Charles~L Lawson and Richard~J Hanson.
\newblock Solving least squares problems.
\newblock 15, 1995.

\bibitem[Hinton et~al.(2006)Hinton, Osindero, and Teh]{hinton2006fast}
Geoffrey~E Hinton, Simon Osindero, and Yee-Whye Teh.
\newblock A fast learning algorithm for deep belief nets.
\newblock \emph{Neural computation}, 18\penalty0 (7):\penalty0 1527--1554,
  2006.

\bibitem[Hinton and Salakhutdinov(2006)]{hinton2006reducing}
Geoffrey~E Hinton and Ruslan~R Salakhutdinov.
\newblock Reducing the dimensionality of data with neural networks.
\newblock \emph{Science}, 313\penalty0 (5786):\penalty0 504--507, 2006.

\bibitem[Hochreiter and Schmidhuber(1997)]{hochreiter1997long}
Sepp Hochreiter and J{\"u}rgen Schmidhuber.
\newblock Long short-term memory.
\newblock \emph{Neural computation}, 9\penalty0 (8):\penalty0 1735--1780, 1997.

\bibitem[Graves(2012)]{graves2012supervised}
Alex Graves.
\newblock Supervised sequence labelling.
\newblock 2012.

\bibitem[Bengio(2009)]{bengio2009learning}
Yoshua Bengio.
\newblock Learning deep architectures for {AI}.
\newblock \emph{Foundations and trends{\textregistered} in Machine Learning},
  2\penalty0 (1):\penalty0 1--127, 2009.

\bibitem[Deng and Kwok(2015)]{dengmirex}
Junqi Deng and Yu-Kwong Kwok.
\newblock {MIREX} 2015 submission: Automatic chord estimation with chord
  correction using neural network, 2015.

\bibitem[Harte(2010)]{harte2010towards}
Christopher Harte.
\newblock \emph{Towards automatic extraction of harmony information from music
  signals}.
\newblock PhD thesis, Department of Electronic Engineering, Queen Mary,
  University of London, 2010.

\bibitem[Cho(2014)]{cho2014improved}
Taemin Cho.
\newblock \emph{Improved techniques for automatic chord recognition from music
  audio signals}.
\newblock PhD thesis, New York University, 2014.

\bibitem[Humphrey(2015)]{humphrey2015exploration}
Eric Humphrey.
\newblock \emph{An Exploration of Deep Learning in Music Informatics.}
\newblock PhD thesis, New York University, 2015.

\bibitem[Srivastava et~al.(2014)Srivastava, Hinton, Krizhevsky, Sutskever, and
  Salakhutdinov]{srivastava2014dropout}
Nitish Srivastava, Geoffrey~E Hinton, Alex Krizhevsky, Ilya Sutskever, and
  Ruslan Salakhutdinov.
\newblock Dropout: a simple way to prevent neural networks from overfitting.
\newblock \emph{Journal of Machine Learning Research}, 15\penalty0
  (1):\penalty0 1929--1958, 2014.

\bibitem[Prechelt(1998)]{prechelt1998automatic}
Lutz Prechelt.
\newblock Automatic early stopping using cross validation: quantifying the
  criteria.
\newblock \emph{Neural Networks}, 11\penalty0 (4):\penalty0 761--767, 1998.

\bibitem[Zeiler(2012)]{zeiler2012adadelta}
Matthew~D Zeiler.
\newblock {ADADELTA}: an adaptive learning rate method.
\newblock \emph{{arXiv} preprint {arXiv}:1212.5701}, 2012.

\bibitem[Geman et~al.(1992)Geman, Bienenstock, and Doursat]{geman1992neural}
Stuart Geman, Elie Bienenstock, and Ren{\'e} Doursat.
\newblock Neural networks and the bias/variance dilemma.
\newblock \emph{Neural computation}, 4\penalty0 (1):\penalty0 1--58, 1992.

\bibitem[Friedman et~al.(2001)Friedman, Hastie, and
  Tibshirani]{friedman2001elements}
Jerome Friedman, Trevor Hastie, and Robert Tibshirani.
\newblock The elements of statistical learning.
\newblock 1, 2001.

\bibitem[Friedman(1937)]{friedman1937use}
Milton Friedman.
\newblock The use of ranks to avoid the assumption of normality implicit in the
  analysis of variance.
\newblock \emph{Journal of the american statistical association}, 32\penalty0
  (200):\penalty0 675--701, 1937.

\bibitem[Tukey(1949)]{tukey1949comparing}
John~W Tukey.
\newblock Comparing individual means in the analysis of variance.
\newblock \emph{Biometrics}, pages 99--114, 1949.

\bibitem[Humphrey and Bello(2015)]{humphreyfour}
Eric~J Humphrey and Juan~P Bello.
\newblock Four timely insights on automatic chord estimation.
\newblock In \emph{Proceedings of the 16th Conference of the International
  Society for Music Information Retrieval (ISMIR)}, 2015.

\bibitem[Koops et~al.(2015)Koops, de~Haas, and Volk]{koopsintegration}
Hendrik~Vincent Koops, W~Bas de~Haas, and Anja Volk.
\newblock Integration of crowd-sourced chord sequences using data fusion.
\newblock In \emph{Proceedings of the 16th International Society for Music
  Information Retrieval Conference, ISMIR}, 2015.

\bibitem[Klein(2004)]{klein2004sensor}
Lawrence~A Klein.
\newblock Sensor and data fusion: a tool for information assessment and
  decision making.
\newblock 324, 2004.

\bibitem[Chawla et~al.(2004)Chawla, Japkowicz, and Kotcz]{chawla2004editorial}
Nitesh~V Chawla, Nathalie Japkowicz, and Aleksander Kotcz.
\newblock Editorial: special issue on learning from imbalanced data sets.
\newblock \emph{ACM Sigkdd Explorations Newsletter}, 6\penalty0 (1):\penalty0
  1--6, 2004.

\end{thebibliography}

\end{document}